\PassOptionsToPackage{unicode}{hyperref}
\PassOptionsToPackage{hyphens}{url}
\PassOptionsToPackage{dvipsnames,svgnames,x11names}{xcolor}
\documentclass[12pt]{article}

\usepackage{iftex}
\ifPDFTeX
  \usepackage[T1]{fontenc}
  \usepackage[utf8]{inputenc}
  \usepackage{textcomp} 
\else 
  \usepackage{unicode-math}
  \defaultfontfeatures{Scale=MatchLowercase}
  \defaultfontfeatures[\rmfamily]{Ligatures=TeX,Scale=1}
\fi
\usepackage{lmodern}
\ifPDFTeX\else  
\fi
\IfFileExists{upquote.sty}{\usepackage{upquote}}{}
\IfFileExists{microtype.sty}{
  \usepackage[]{microtype}
  \UseMicrotypeSet[protrusion]{basicmath} 
}{}
\makeatletter
\@ifundefined{KOMAClassName}{
  \IfFileExists{parskip.sty}{%
    \usepackage{parskip}
  }{
    \setlength{\parindent}{0pt}
    \setlength{\parskip}{6pt plus 2pt minus 1pt}}
}{
  \KOMAoptions{parskip=half}}
\makeatother
\usepackage{xcolor}
\makeatletter
\ifx\paragraph\undefined\else
  \let\oldparagraph\paragraph
  \renewcommand{\paragraph}{
    \@ifstar
      \xxxParagraphStar
      \xxxParagraphNoStar
  }
  \newcommand{\xxxParagraphStar}[1]{\oldparagraph*{#1}\mbox{}}
  \newcommand{\xxxParagraphNoStar}[1]{\oldparagraph{#1}\mbox{}}
\fi
\ifx\subparagraph\undefined\else
  \let\oldsubparagraph\subparagraph
  \renewcommand{\subparagraph}{
    \@ifstar
      \xxxSubParagraphStar
      \xxxSubParagraphNoStar
  }
  \newcommand{\xxxSubParagraphStar}[1]{\oldsubparagraph*{#1}\mbox{}}
  \newcommand{\xxxSubParagraphNoStar}[1]{\oldsubparagraph{#1}\mbox{}}
\fi
\makeatother

\usepackage{longtable,booktabs,array}
\usepackage{calc} 
\usepackage{etoolbox}
\makeatletter
\patchcmd\longtable{\par}{\if@noskipsec\mbox{}\fi\par}{}{}
\makeatother
\IfFileExists{footnotehyper.sty}{\usepackage{footnotehyper}}{\usepackage{footnote}}
\makesavenoteenv{longtable}
\usepackage{graphicx}
\makeatletter
\def\maxwidth{\ifdim\Gin@nat@width>\linewidth\linewidth\else\Gin@nat@width\fi}
\def\maxheight{\ifdim\Gin@nat@height>\textheight\textheight\else\Gin@nat@height\fi}
\makeatother
\setkeys{Gin}{width=\maxwidth,height=\maxheight,keepaspectratio}
\makeatletter
\def\fps@figure{htbp}
\makeatother

\makeatletter
\@ifpackageloaded{caption}{}{\usepackage{caption}}
\AtBeginDocument{%
\ifdefined\contentsname
  \renewcommand*\contentsname{Table of contents}
\else
  \newcommand\contentsname{Table of contents}
\fi
\ifdefined\listfigurename
  \renewcommand*\listfigurename{List of Figures}
\else
  \newcommand\listfigurename{List of Figures}
\fi
\ifdefined\listtablename
  \renewcommand*\listtablename{List of Tables}
\else
  \newcommand\listtablename{List of Tables}
\fi
\ifdefined\figurename
  \renewcommand*\figurename{Figure}
\else
  \newcommand\figurename{Figure}
\fi
\ifdefined\tablename
  \renewcommand*\tablename{Table}
\else
  \newcommand\tablename{Table}
\fi
}
\@ifpackageloaded{float}{}{\usepackage{float}}
\floatstyle{ruled}
\@ifundefined{c@chapter}{\newfloat{codelisting}{h}{lop}}{\newfloat{codelisting}{h}{lop}[chapter]}
\floatname{codelisting}{Listing}

\makeatother
\makeatletter
\@ifpackageloaded{caption}{}{\usepackage{caption}}
\@ifpackageloaded{subcaption}{}{\usepackage{subcaption}}
\makeatother

\ifLuaTeX
  \usepackage{selnolig}  
\fi
\usepackage[]{natbib}
\usepackage{bookmark}

\IfFileExists{xurl.sty}{\usepackage{xurl}}{} 
\usepackage{orcidlink}

\usepackage{amsmath,amssymb}
\usepackage{amsthm}
\usepackage{bm}
\usepackage{bbm}
\usepackage{mathrsfs}
\usepackage{enumitem}
\usepackage{multirow}

\usepackage[linesnumbered,ruled]{algorithm2e}
\SetKwRepeat{Do}{do}{while}

\theoremstyle{plain}
\newtheorem{theorem}{Theorem}
\newtheorem{lemma}{Lemma}
\newtheorem{proposition}{Proposition}

\theoremstyle{remark}

\theoremstyle{definition}

\newtheorem{assumption}{Assumption}

\usepackage{xr-hyper}
\usepackage{hyperref}
\makeatletter
\newcommand*{\addFileDependency}[1]{
\typeout{(#1)}
\@addtofilelist{#1}
\IfFileExists{#1}{}{\typeout{No file #1.}}
}
\makeatother

\hypersetup{
  bookmarksnumbered=true,
  colorlinks=true,
  citecolor=ForestGreen,       
  linkcolor=Blue, 		     
  urlcolor=MidnightBlue     	 
}

\graphicspath{{../figure/}}

\newcommand{\E}{\mathbb{E}}
\newcommand{\N}{\mathcal{N}}
\newcommand{\1}{\mathbbm{1}}

\newcommand{\argmax}{\operatornamewithlimits{arg\,max}}

\newcommand{\Cov}{\operatorname{Cov}}
\newcommand{\Var}{\operatorname{Var}}

\usepackage{etoolbox}
\newcommand{\renewSformat}[1]{%
  \expandafter\renewcommand\csname the#1\endcsname{S\arabic{#1}}%
  \setcounter{#1}{0}%
}

\newcommand{\anon}{1}

\begin{document}

\def\spacingset#1{\renewcommand{\baselinestretch}%
{#1}\small\normalsize} \spacingset{1}

\if1\anon
{
  \title{\bf Structural Analysis of Object-Valued Time Series}
  \author{Mingyuan Xu\thanks{Xu and Huang contributed equally to this work.}~\orcidlink{0009-0000-3899-4205},
\quad
Rundong Huang\footnotemark[1]~\orcidlink{0009-0008-3369-6987},
\quad
Doudou Zhou\thanks{Corresponding authors:
\href{mailto:ddzhou@nus.edu.sg}{\texttt{ddzhou@nus.edu.sg}}
and
\href{mailto:tongyuli@nus.edu.sg}{\texttt{tongyuli@nus.edu.sg}}.}~\orcidlink{0000-0002-0830-2287},
\quad
Tongyu Li\footnotemark[2]~\orcidlink{0009-0007-5051-2716}
\\[0.5em]
Department of Statistics and Data Science,
National University of Singapore
}
    \date{}
  \maketitle
} \fi

\if0\anon
{
  \bigskip
  \bigskip
  \bigskip
  \begin{center}
    {\LARGE\bf Structural Analysis of Object-Valued Time Series}
\end{center}
  \medskip
} \fi

\bigskip
\begin{abstract}
We develop inference for structural patterns in the temporal association of object-valued time series. Our framework is built on pairwise lag profiles, which record the expected value of a chosen pairwise function at each lag and serve as a unified inferential target. We consider three representative structural features: whether a profile reaches stability beyond a finite lag, remains constant over time, and displays geometric regularity at short lags. To assess these properties, we design statistical tests calibrated by a common block multiplier bootstrap that accommodates serial dependence and the overlap inherent in lagged pairs. Gaussian approximation results establish bootstrap validity for asymptotic size control, while suitable signal conditions guarantee consistency against the corresponding alternatives. Finite-sample performance is examined through simulations in both Euclidean and density-valued settings. Applications to electricity load curves and stock-return distributions illustrate how the proposed procedures characterize association across lags.
\end{abstract}

\noindent%
\vfill

\newpage
\spacingset{1.8} 


\section{Introduction}
\label{sec:intro}
Modern time-series applications increasingly involve object-valued observations, such as functions \citep{hormann2010weakly}, covariance matrices \citep{golosnoy2012conditional}, networks \citep{dubey2022modeling}, and probability distributions \citep{zhang2022Wasserstein}. In these settings, scientific interest often extends beyond detecting serial dependence to examining how temporal association is structured. For example, one may ask over what range of lags temporal association varies, whether it remains unchanged over the observation period, and whether it follows a simple parametric form. Analogous structural questions have been studied for scalar and multivariate data, including tests for banded dependence among ordered variables \citep{qiu2012test, yao2018testing}, changes in dependence over time \citep{cho2015multiple, dette2019change, mcgonigle2025nonparametric}, and parsimonious models for correlation as a function of lag \citep{simpson2010linear}. Addressing these questions for general random objects requires a notion of temporal association that is both broadly applicable across object spaces and amenable to statistical inference.

Developing such a notion is nontrivial because general object spaces may lack local or global linear structure, making classical methods based on vector operations and inner products inapplicable \citep{dubey2020functional}. This challenge has motivated the development of approaches based on Fr\'echet summaries \citep{dubey2019frechet} and pairwise constructions built on distances and kernel similarities \citep{szekely2007measuring,sejdinovic2013equivalence,dubey2024metric}. In the time-series setting, existing methods have primarily focused on detecting serial dependence, either at fixed lags or through omnibus aggregation across lags \citep{jiang2024testing,zhu2025wise,li2026autohsic,zhang2026auto}, while more general procedures target the full i.i.d.\ hypothesis \citep{li2025general}. Although these methods provide powerful tools for identifying dependence, they offer limited insight into the structure of temporal association itself. Closer in spirit to our work, \citet{seo2026longrange} investigated long-range dependence in object-valued time series using trace autocovariances, which admit a distance-based representation. Their focus is on the asymptotic decay of dependence as the lag diverges, whereas our interest lies in characterizing the diverse behavior of temporal association and in understanding how its structure evolves over time.

Existing work on structural features of temporal dependence remains fragmented and largely confined to specific object geometries. For functional time series, changes in second-order structure have been studied through autocovariance operators: \citet{zhang2011testing} considered the lag-one operator, and \citet{bucher2020detecting} examined operators across multiple lags as part of a test for second-order stationarity. These approaches rely on the Hilbert-space structure of functional data. For general random objects, change-point procedures have targeted changes in Fr\'echet summaries \citep{dubey2020frechet,2024Jiangtwo} or in marginal distributions \citep{dubey2026change,zhang2026change}. Yet temporal association may change even when the marginal distribution remains unchanged, and such changes have received comparatively little attention in general object spaces. 
At the modeling level, autoregressive models have been developed in Wasserstein spaces \citep{zhang2022Wasserstein,zhu2023autoregressive,ghodrati2024distributional}, on spheres \citep{zhu2024spherical}, and in Hadamard spaces \citep{bulte2024autoregressive}. While these models describe the dynamics of the object-valued process itself, they do not directly assess whether temporal association exhibits a simple parametric form. 
Overall, there is still a lack of a unified inferential framework for studying structural features of temporal association across general objects.

To provide such a framework, we build on the classical lag-indexed view of serial dependence. Autocovariances summarize linear dependence across lags \citep{brockwell2009time}; distance-based lag measures extend this perspective to nonlinear dependence \citep{zhou2012measuring,davis2018applications}, and distance- and kernel-based measures extend it further to object-valued time series \citep{jiang2024testing,li2026autohsic,seo2026longrange}. We represent the aspect of temporal association under investigation through a newly introduced \emph{pairwise lag profile}. For a chosen pairwise function, the profile records how its expectation varies with temporal separation. Distances, similarities, and other scalar-valued pairwise functions therefore induce different, interpretable targets for structural inference. 

Within this framework, structural features of temporal association can be expressed as restrictions on the pairwise lag profile and studied through hypothesis testing. We develop inference for three representative hypotheses. 
The first, \emph{profile stabilization}, asks whether the profile becomes constant beyond a cutoff, up to a prespecified lag horizon. Unlike the autocorrelation cutoff property of a moving-average process, the resulting plateau need not occur at zero. Rather, stabilization indicates that additional temporal separation beyond the cutoff produces no systematic change in the profile.
The second, \emph{temporal stability}, examines whether the profile remains unchanged over the observation period, with alternatives allowing a change at an unknown time point. 
The third, \emph{geometric regularity}, concerns whether the initial segment of the profile satisfies the curvature identities implied by a single geometric sequence. We employ this restriction as a parsimonious benchmark for short-lag structure without imposing an autoregressive recursion on the object-valued process itself. 

All the three hypotheses can be analyzed through a common empirical construction. Up to a chosen lag horizon, we collect pairwise scores into lag-score vectors, whose time average estimates the population lag profile. Profile stabilization, temporal stability, and geometric regularity are then assessed through cumulative contrasts, sequential bridge statistics, and a curvature transformation, respectively. Because each coordinate of the lag-score vectors depends on pairs of observations separated by the corresponding lag, the lag horizon governs both their dimension and the range of their dependence. To characterize the null distributions of the resulting test statistics, we derive moment and dependence bounds that are explicit in the lag horizon and establish Gaussian approximations through finite-memory approximation and blocking, using a high-dimensional central limit theorem for independent sums \citep{chernozhukov2017central}. For calibration, we use a block Gaussian multiplier process \citep{zhangcheng2014bootstrapping,yu2021finite} and quantify how much of the correlation induced by overlapping pairs its blocks fail to capture. The block length required for accurate calibration thus depends on the lag horizon, not only on the serial dependence of the underlying series. Applying the task-specific transformations to this multiplier process yields bootstrap counterparts for all three test statistics. 

Our contributions can be summarized in the following four parts. 
First, we formulate the pairwise lag profile as a unified inferential target for structural analysis of temporal association. Because its definition requires no linear structure, it applies to a broad class of object spaces and does not rely on a generative model for the object-valued process. 
Second, we develop tests for three representative structural properties of the profile, namely stabilization, temporal stability, and geometric regularity. Each test is built from a task-specific functional of the lag-score process, and a common block multiplier bootstrap is employed to calibrate it. 
Third, we establish Gaussian and bootstrap approximation results that account for the dual role of the lag horizon as both a dimensional and dependence parameter, as well as for the additional correlation induced by overlapping pairs. These results yield theoretical guarantees for asymptotically valid inference across the three testing problems. 
Finally, we complement the developments with simulations in Euclidean and density-valued settings to evaluate finite-sample size and power. Applications to electricity load curves and cross-sectional stock-return distributions further illustrate the scope and interpretability of the proposed framework.

The rest of this paper is organized as follows. Section~\ref{sec:method} introduces the pairwise lag-profile framework and the proposed testing procedures. Section~\ref{sec:theory} presents the corresponding theoretical guarantees. Sections~\ref{sec:simulation} and~\ref{sec:real_data} report the simulation studies and real-data applications, respectively. Technical proofs and auxiliary results are provided in the Supplementary Material.

\section{Methodology}
\label{sec:method}

\subsection{Pairwise lag profiles}
\label{subsec:method-profile}

Let $(\mathcal X,d)$ be a metric space and let $\{X_t\}_{t\in\mathbb Z}$ be an $\mathcal X$-valued time series.  The metric-space formulation accommodates objects such as functions, distributions, networks, and other non-Euclidean observations without requiring addition, multiplication, or an inner product on $\mathcal X$.  To extract a meaningful scalar feature from a pair of observations, we work with a measurable pairwise function
\[
  h:\mathcal X\times\mathcal X\to\mathbb R
\]
chosen according to the feature of serial structure under study.  
The role of $h$ is analogous to choosing which aspect of dependence is to be summarized. Table~\ref{tab:pairwise-function-examples} in the Supplementary Material summarizes several representative choices of the pairwise function and illustrates how each choice determines the interpretation of the resulting lag profile.

Once $h$ is fixed, we index its pairwise summary by lag, as in the classical autocovariance function for scalar time series \citep{brockwell2009time}. Lag-specific dependence measures have also been developed for object-valued series, although they need not measure the same feature \citep{jiang2024testing,li2025general}. On a stationary segment, the population quantity associated with lag $k$ is
\begin{equation}
  \theta_k:=\E\{h(X_0,X_k)\},\qquad k\ge1,
  \label{eq:common-theta}
\end{equation}
whenever the expectation exists. We call the sequence $(\theta_k)_{k\ge1}$ the \emph{pairwise lag profile}, which describes the feature encoded by $h$.

The population profile \eqref{eq:common-theta} is not observed directly. To estimate several lags jointly, we collect all pairwise scores anchored at a time point into a single vector. Let \(K_n\) denote the lag horizon, that is, the largest lag considered, and set \(N_n=n-K_n\). For \(1\le t\le N_n\), the resulting lag-score vector is
\begin{equation}
H_{t,n}:=\bigl(h(X_t,X_{t+1}),\ldots,h(X_t,X_{t+K_n})\bigr)^\top
  \in\mathbb R^{K_n}.
  \label{eq:lag-score-vector}
\end{equation}
The empirical counterpart of the population profile up to the lag horizon, \(\Theta_n:=(\theta_1,\ldots,\theta_{K_n})^\top\), is given by averaging the lag-score vectors, that is,
\begin{equation}
  \widehat\Theta_n:=\frac{1}{N_n}\sum_{t=1}^{N_n}H_{t,n},
  \qquad
  \widehat\theta_{k,n}:=\frac{1}{N_n}\sum_{t=1}^{N_n}h(X_t,X_{t+k}).
  \label{eq:common-denom-estimator}
\end{equation}
The common anchor range ensures that every lag coordinate is estimated from the same number of pairs. 

For joint inference on the lag profile, consider the estimation error as a normalized sum of dependent vectors,
\[
  U_n:=\sqrt{N_n}(\widehat\Theta_n-\Theta_n)
  =N_n^{-1/2}\sum_{t=1}^{N_n}(H_{t,n}-\Theta_n).
\]
Stabilization and geometric regularity will be analyzed through linear contrasts and a linearization of this vector.
To assess temporal stability, we compare each partial sum of the lag-score vectors with the corresponding fraction of the full-sample sum, following the CUSUM approach to change-point detection \citep{zhang2011testing,yu2021finite}. 
For \(s\in[0,1]\), define
\begin{equation}
  \begin{aligned}
    B_n(s)
    &:=N_n^{-1/2}\left\{
      \sum_{t=1}^{\lfloor N_ns\rfloor}H_{t,n}
      -\frac{\lfloor N_ns\rfloor}{N_n}\sum_{t=1}^{N_n}H_{t,n}
    \right\}=N_n^{-1/2}\sum_{t=1}^{\lfloor N_ns\rfloor}
      (H_{t,n}-\widehat\Theta_n).
  \end{aligned}
  \label{eq:observable-bridge}
\end{equation}
Note that \(B_n(0)=B_n(1)=0\). Under temporal stability, \(B_n(s)\) fluctuates around zero, whereas a change in the lag profile produces a systematic drift around the break location.

\subsection{Tests for structural properties}
\label{subsec:method-structural}
We now formulate three structural properties of temporal association as restrictions on the population pairwise lag profile, and construct their tests from $U_n$ and $B_n$.  Their difference lies in the restriction being tested and in the transformation applied to the terminal value or bridge, summarized in Table~\ref{tab:structural-restrictions} in the Supplementary Material.

\subsubsection{Profile stabilization}
\label{subsec:method-stabilization}

The first task concerns whether the lag profile is constant beyond a prespecified cutoff, up to the lag horizon, where the common plateau level is unrestricted and may be nonzero. 

For a candidate cutoff \(q_0\in\{0,1,\ldots,K_n-2\}\), let \(m_{q_0}:=K_n-q_0\) and define the tail-selection matrix \( P_{q_0} := \begin{pmatrix} 0_{m_{q_0}\times q_0} & I_{m_{q_0}} \end{pmatrix}. \) Then \( P_{q_0}\Theta_n = (\theta_{q_0+1},\ldots,\theta_{K_n})^\top \) is the part of the lag profile used to assess stabilization beyond \(q_0\).
Stabilization means that the tail coordinates share an exact unknown level. A tail that approaches a limiting level but remains nonconstant, as can occur for autoregressive processes, belongs to the alternative. The null and alternative can therefore be stated without introducing that nuisance level explicitly:
\[
  \begin{aligned}
    H_0^{\mathrm{stab}}(q_0): &\quad P_{q_0}\Theta_n \in \operatorname{span}(\1_{m_{q_0}}), \\
    H_a^{\mathrm{stab}}(q_0): &\quad P_{q_0}\Theta_n \notin \operatorname{span}(\1_{m_{q_0}}).
  \end{aligned}
\]
The unknown plateau level can be removed by centering the tail profile. Cumulative sums then aggregate departures along the ordered lag coordinates, and their largest absolute value measures the departure from a constant tail. These two operations are represented by \( H_{m_{q_0}}:=I_{m_{q_0}}-m_{q_0}^{-1}\1_{m_{q_0}}\1_{m_{q_0}}^\top, \   C_{m_{q_0}} := \bigl(\1\{j\le i\}\bigr)_{1\le i\le m_{q_0}-1,\,1\le j\le m_{q_0}}. \)
A departure from stabilization is therefore measured by the largest absolute cumulative contrast.  Scaling by the estimation rate gives the test statistic
\[ 
R_n(q_0) := \sqrt{N_n}\, \bigl\| C_{m_{q_0}}H_{m_{q_0}} P_{q_0}\widehat\Theta_n \bigr\|_\infty. 
\]

Under \(H_0^{\mathrm{stab}}(q_0)\), the population tail is removed exactly, so \(R_n(q_0) = \bigl\| C_{m_{q_0}}H_{m_{q_0}} P_{q_0} U_{n} \bigr\|_\infty\) measures the fluctuations of a linear contrast of \(U_n\).
Applying the same contrast to \(U_n^*\), defined in \eqref{eq:bootstrap-partial-sum} in Section~\ref{subsec:method-implementation}, leads to the bootstrap statistic
\[
  R_n^*(q_0) := \bigl\|C_{m_{q_0}}H_{m_{q_0}}P_{q_0} U_{n}^*\bigr\|_\infty.
\]
Let \(\widehat c_{n,1-\alpha}^{\mathrm{stab}}(q_0)\) be the conditional \((1-\alpha)\)-quantile of \(R_n^*(q_0)\). For a prespecified cutoff $q_0$, we reject $H_0^{\mathrm{stab}}(q_0)$ when $R_n(q_0)>\widehat c_{n,1-\alpha}^{\mathrm{stab}}(q_0)$.
For a fixed lag horizon $K$, we estimate the first stabilization cutoff by the first candidate whose null hypothesis is not rejected:
\[
  \widehat q_{n,\alpha}:=\min\left\{q_0\in\{0,\ldots,K-2\}:R_n(q_0)\le\widehat c_{n,1-\alpha}^{\mathrm{stab}}(q_0)\right\},
\]
with $\min\varnothing=\infty$. The value $\widehat q_{n,\alpha}=\infty$ indicates that no plateau is accepted among the candidate cutoffs.

\subsubsection{Temporal stability}
\label{subsec:method-cp}

The second task concerns whether the selected lag profile remains unchanged over calendar time. Its null hypothesis requires the lag-wise pairwise expectations to remain constant throughout the observation period. 

Since \(\E H_{t,n}\) collects the lag-wise pairwise means at anchor time \(t\), temporal stability can be formulated as
\[
  \begin{aligned}
    H_0^{\mathrm{temp}}: &\quad \E H_{t,n}=\Theta_n \quad\text{for all }1\le t\le N_n, \\
    H_a^{\mathrm{temp}}: &\quad \E H_{t_1,n}\ne\E H_{t_2,n} \quad\text{for some }t_1,t_2.
  \end{aligned}
\]
The null requires the selected lag profile to remain constant over time, while the alternative allows at least one lag coordinate to change. The asymptotic validity of the test is established for stationary processes satisfying Assumptions~\ref{asm:data-score}--\ref{asm:block-complexity} in Section~\ref{sec:theory}; detection and localization are studied under a single change between stationary regimes.

Under $H_0^{\mathrm{temp}}$, the observable bridge $B_n$ in \eqref{eq:observable-bridge} fluctuates around zero.  A change at an unknown time can affect an unknown subset of lag coordinates, so the test should search simultaneously over both indices. 
We restrict candidate change points to an interior region so that both sides of each split contain a nonvanishing fraction of the observations. For a fixed $\epsilon\in(0,1/2)$, the admissible change fractions are
\[
  \mathcal S_n(\epsilon) := \{m/N_n:m\in\mathbb Z,\ \lfloor\epsilon N_n\rfloor\le m\le\lfloor(1-\epsilon)N_n\rfloor\}.
\]
Maximizing over candidate change points and lag coordinates gives the test statistic 
\[
  T_n(K_n,\epsilon) := \sup_{s\in\mathcal S_n(\epsilon)} \|B_n(s)\|_\infty.
\]

The bootstrap statistic applies the same lag--time maximum to \(B_n^*(s)\), defined in \eqref{eq:bootstrap-partial-sum-bridge} in Section~\ref{subsec:method-implementation}, and is given by 
\[
  T_n^*(K_n,\epsilon) := \sup_{s\in\mathcal S_n(\epsilon)} \|B_n^*(s)\|_\infty.
\]
Let $\widehat c_{n,1-\alpha}^{\mathrm{CP}}$ denote the conditional $(1-\alpha)$-quantile of $T_n^*(K_n,\epsilon)$.  We reject $H_0^{\mathrm{temp}}$ when \( T_n(K_n,\epsilon) > \widehat c_{n,1-\alpha}^{\mathrm{CP}}. \)
When the null is rejected, we consider the change location under a single-change model with stationary regimes on either side of the break. We estimate the break fraction by
\[
  \widehat\tau_n \in \argmax_{s\in\mathcal S_n(\epsilon)} \|B_n(s)\|_\infty.
\]

\subsubsection{Geometric regularity}
\label{subsec:method-ar}

The third task concerns whether the short-lag profile satisfies the local curvature restrictions implied by a geometric sequence. Geometric decay provides a benchmark familiar from scalar autoregressive models, without imposing an autoregressive recursion on the object-valued observations themselves. 

The motivating geometric benchmark is
\[
  H_0^{\mathrm{geo}}:
  \exists\ c\ne0,\ \rho\in(0,1)
  \text{ such that }\theta_k=c\rho^{k-1},
  \qquad 1\le k\le K_n.
\]
Every such sequence satisfies \(\theta_k\theta_{k+2}-\theta_{k+1}^2=0\) for \(1\le k\le K_n-2\). To test these restrictions without estimating \(c\) or \(\rho\), define  \(\Phi_n(\theta):=\bigl(\theta_1\theta_3-\theta_2^2,\ldots,\theta_{K_n-2}\theta_{K_n}-\theta_{K_n-1}^2\bigr)^\top.\)
The null and alternative can then be stated as
\[
  \begin{aligned}
    H_0^{\mathrm{curv}}: &\quad \Phi_n(\Theta_n)=0, \\
    H_a^{\mathrm{curv}}: &\quad \Phi_n(\Theta_n)\ne0.
  \end{aligned}
\]
Thus the null requires every consecutive triple to satisfy the curvature identity, while the alternative allows at least one violation. These identities do not impose \(\rho\in(0,1)\) or exclude degenerate profiles, so \(H_0^{\mathrm{curv}}\) is weaker than the geometric benchmark \(H_0^{\mathrm{geo}}\).

The empirical discrepancy is obtained by evaluating the same map at $\widehat\Theta_n$ and scaling it at the root-$N_n$ rate, so the resulting test statistic is
\[
  A_n(K_n) := \sqrt{N_n}\, \|\Phi_{n}(\widehat\Theta_n)\|_\infty.
\]
Its first-order fluctuation is determined by the Jacobian $D\Phi_{n}(\Theta_n)$. Writing \(\phi_k\) for the \(k\)th component of \(\Phi_n\), the nonzero derivatives are \( \frac{\partial\phi_k}{\partial\theta_k}=\theta_{k+2}, \, \frac{\partial\phi_k}{\partial\theta_{k+1}}=-2\theta_{k+1}, \, \frac{\partial\phi_k}{\partial\theta_{k+2}}=\theta_k. \) 
We estimate this Jacobian at \(\widehat\Theta_n\) and apply it to the bootstrap vector \(U_n^*\) to obtain 
\[
  A_n^*(K_n) := \bigl\| D\Phi_{n}(\widehat\Theta_n) U_n^* \bigr\|_\infty.
\]
Let $\widehat c_{n,1-\alpha}^{\mathrm{curv}}(K_n)$ denote the conditional $(1-\alpha)$-quantile of $A_n^*(K_n)$.  The procedure then rejects $H_0^{\mathrm{curv}}$ when \( A_n(K_n)>\widehat c_{n,1-\alpha}^{\mathrm{curv}}(K_n). \) 

\subsection{Block Gaussian bootstrap calibration}
\label{subsec:method-implementation}
The lag-score sequence inherits serial dependence from the observations and additional dependence from overlapping pairs. We account for this dependence using Gaussian multipliers that are constant within blocks.

For a block length \(b_n\), let \(L_n:=\lceil N_n/b_n\rceil\) and partition \(\{1,\ldots,N_n\}\) into
\begin{equation}
  I_{\ell,n}:=\{(\ell-1)b_n+1,\ldots,\min(\ell b_n,N_n)\},
  \qquad \ell=1,\ldots,L_n.
  \label{eq:block-partition}
\end{equation}

For each bootstrap replication, generate \(\xi_1,\ldots,\xi_{L_n}\stackrel{\mathrm{i.i.d.}}{\sim}\N(0,1)\) independently of the data. Following the blockwise multiplier approach of \citet{zhangcheng2014bootstrapping}, define the bootstrap counterpart of \(U_n=\sqrt{N_n}(\widehat\Theta_n-\Theta_n)\) by
\begin{equation}
  U_n^*
  :=N_n^{-1/2}\sum_{\ell=1}^{L_n}\xi_\ell
  \sum_{t\in I_{\ell,n}}(H_{t,n}-\widehat\Theta_n).
  \label{eq:bootstrap-partial-sum}
\end{equation}
This vector is used to calibrate the profile-stabilization and geometric-regularity tests.
For temporal stability, apply the CUSUM weights to the centered lag-score vectors, using the same block multipliers throughout the scan. Following the block-multiplier CUSUM principle discussed in \citet{yu2021finite}, define the bootstrap process of \eqref{eq:observable-bridge} by 
\begin{equation}
  B_n^*(s)
  :=N_n^{-1/2}\sum_{\ell=1}^{L_n}\xi_\ell
  \sum_{t\in I_{\ell,n}}
  \left\{
    \1\{t\le\lfloor N_ns\rfloor\}
    -\frac{\lfloor N_ns\rfloor}{N_n}
  \right\}
  (H_{t,n}-\widehat\Theta_n),
  \qquad s\in[0,1].
  \label{eq:bootstrap-partial-sum-bridge}
\end{equation}


In the numerical studies, we take \(b_n=\lceil cK_n^{2/3}N_n^{1/3}\rceil\) with \(c=0.3\); Section~\ref{subsec:method-tuning} in the Supplementary Material discusses this choice, and Section~\ref{supp:block-sensitivity} examines its sensitivity. Algorithm~\ref{alg:lag-profile-implementation} in the Supplementary Material summarizes the computation.

\section{Theoretical Results}
\label{sec:theory}

We establish Gaussian and bootstrap approximations for the lag-score process and then derive size and power guarantees for the three testing procedures. We also characterize the selection of the first plateau cutoff and establish detection and localization guarantees under a single temporal change. The profile-stabilization and geometric-regularity tests use a fixed lag horizon, while the temporal-stability test allows $K_n$ to grow under suitable dependence and approximation conditions.


\subsection{Dependence assumptions for lag-score processes}
\label{subsec:theory-conditions}

We use the physical-dependence framework of \citet{wu2005nonlinear}, which measures the effect of replacing one innovation in a causal representation. This framework also underlies Gaussian approximation for high-dimensional time series \citep{zhangwu2017gaussian}. Assumption~\ref{asm:data-score} combines a polynomial physical-dependence bound with metric moment and growth-Lipschitz conditions that transfer dependence from the observations to their pairwise scores; see Lemma~\ref{lem:dependence-transfer} in the Supplementary Material.

\begin{assumption}[Data and score regularity]
\label{asm:data-score}
The process under study has a causal Bernoulli-shift representation \(X_t=g(\ldots,\varepsilon_{t-1},\varepsilon_t)\), where \(\{\varepsilon_t\}\) are i.i.d. innovations in a standard Borel space. Let \(X_j^*\) be obtained by replacing \(\varepsilon_0\) with an independent copy, and define \(\delta_{j,u}^X:=\|d(X_j,X_j^*)\|_u\) for \(u\ge1\). There exist \(q\ge4\), \(\beta>2\), \(\gamma\ge0\), and \(x_0\in\mathcal X\) such that:
\begin{enumerate}
\item[(i)] For some \(C_X<\infty\), \(\delta_{j,q(1+\gamma)}^X\le C_X(j+1)^{-\beta}\) for every \(j\ge0\).
\item[(ii)] For some \(L_h<\infty\) and all \(x_1,x_2,y_1,y_2\in\mathcal X\),
\[
\begin{aligned}
|h(x_1,y_1)-h(x_2,y_2)|&\le L_h\{1+d(x_1,x_0)+d(y_1,x_0)+d(x_2,x_0)+d(y_2,x_0)\}^{\gamma}\\
&\qquad\times\{d(x_1,x_2)+d(y_1,y_2)\}.
\end{aligned}
\]
\item[(iii)] The metric moment satisfies \(\|d(X_0,x_0)\|_{q(1+\gamma)}<\infty\).
\end{enumerate}
\end{assumption}

The innovation coupling modifies \cite{wu2005nonlinear} so that the metric formulation and the regularity of $h$ make it applicable to the present object-valued observations. The parameter \(\gamma\) allows the sensitivity of \(h\) to grow with the magnitude of its arguments. Distance-based pairwise functions correspond to \(\gamma=0\), while inner products are covered by \(\gamma=1\) with stronger observation moments. Table~\ref{tab:score-regularity} in the Supplementary Material records the corresponding sufficient conditions.

The long-run covariance matrix of the lag-score process is
\begin{equation}
  \Sigma_n:=\sum_{r\in\mathbb Z}\Cov(H_{0,n},H_{r,n}).
  \label{eq:common-long-run-covariance}
\end{equation}
Under Assumption~\ref{asm:data-score}, Lemma~\ref{lem:app-covariance-summability} in the Supplementary Material establishes entrywise absolute convergence of the series defining \(\Sigma_n\) and the uniform bound \(\sup_n\max_{1\le k\le K_n}(\Sigma_n)_{kk}<\infty\).
Gaussian approximation for maxima additionally requires nondegeneracy in the directions entering the statistic \citep{chernozhukov2015comparison}. To specify these directions, let \(\mathcal T_n\) consist of the rows of the stabilization contrast matrices \(C_{m_{q_0}}H_{m_{q_0}}P_{q_0}\), the coordinate directions used in the temporal scan, and the rows of \(D\Phi_n(\Theta_n)\) for geometric regularity, together with their negatives, retaining only those relevant to the procedure under consideration. We impose the following lower variance bound on these directions.
\begin{assumption}[Transformed nondegeneracy]
\label{asm:covariance-nondegeneracy}
There exists a constant \(\underline\sigma>0\), independent of \(n\), such that \(\inf_n\inf_{a\in\mathcal T_n}a^\top\Sigma_na\ge\underline\sigma^2\).
\end{assumption}
For temporal inference, the variance lower bound in Assumption~\ref{asm:covariance-nondegeneracy}, together with trimming, keeps the Gaussian bridge variances bounded away from zero over the scan. For geometric regularity, it ensures nondegeneracy in the Jacobian directions and thus excludes profiles such as the all-zero profile, where the first-order approximation degenerates.

We establish Gaussian approximation by replacing the centered scores with finite-memory approximations and separating independent large blocks, to which we apply \citet[Proposition~2.1]{chernozhukov2017central}. Bootstrap calibration additionally requires control of the block-multiplier covariance \citep{zhangcheng2014bootstrapping,chernozhukov2015comparison}; we therefore separate its population block-covariance bias from its sampling error. Let \(\|\cdot\|_{\max}\) denote the largest absolute matrix entry and define
\[
  \mathfrak b_n:=\left\|\frac{1}{b_n}\Cov\left(\sum_{t=1}^{b_n}H_{t,n}\right)-\Sigma_n\right\|_{\max}.
\]
This quantity measures the discrepancy between the covariance captured by a block and the long-run covariance. Let $d_n^{\max}$ denote the largest number of signed coordinates in the Gaussian maxima, and set $\ell_n:=\log(2N_nK_n+2d_n^{\max})$. Write $a\vee b:=\max\{a,b\}$.

\begin{assumption}[Block-complexity conditions]
\label{asm:block-complexity}
The following conditions hold:
\begin{enumerate}
\item[(i)] There exist positive integer sequences $m_n$ and $r_n$ such that, with $M_n:=K_n+m_n$,
\[
K_n^{1/q}m_n^{1-\beta}\ell_n^2\to0,\qquad
K_n^{2/q}\frac{M_n}{r_n}\ell_n^2\to0,\qquad
K_n^{2/q}\left(\frac{r_n}{N_n}\right)^{1-2/q}\ell_n^7\to0.
\]
\item[(ii)] The bootstrap block length satisfies $b_n\to\infty$ and $b_n/N_n\to0$, and
\[
\left\{\mathfrak b_n+K_n^{4/q}\left(\frac{b_n}{N_n}\right)^{1/2}\right\}\{1\vee\log(2d_n^{\max})\}^2\to0.
\]
\end{enumerate}
\end{assumption}

In Assumption~\ref{asm:block-complexity}(i), $m_n$ and $r_n$ are auxiliary truncation and block lengths; the three limits control truncation error, separating gaps, and Gaussian approximation of the independent block sums. Assumption~\ref{asm:block-complexity}(ii) controls the bias and sampling error of the bootstrap covariance. For fixed-window terminal procedures, $b_n\to\infty$ and $b_n=o(N_n)$ suffice for part~(ii), while temporal scans additionally require control of the logarithmic factor, even for fixed $K$.


\subsection{Gaussian and bootstrap approximations}
\label{subsec:theory-master}

Theorems~\ref{thm:master-gaussian} and \ref{thm:master-bootstrap} below establish Gaussian and conditional block-bootstrap approximations for terminal and bridge maxima. The terminal results yield size validity for stabilization and, after linearization, geometric regularity in Sections~\ref{subsec:theory-stabilization} and \ref{subsec:theory-ar}, while the bridge results calibrate the temporal test in Section~\ref{subsec:theory-cp}.

To define the Gaussian targets, let \(Z_{1,n},\ldots,Z_{N_n,n}\overset{\mathrm{i.i.d.}}{\sim}\N(0,\Sigma_n)\) and set
\[
G_n(s):=N_n^{-1/2}\sum_{t=1}^{\lfloor N_ns\rfloor}Z_{t,n},\qquad G_n^{\mathrm{br}}(s):=G_n(s)-\frac{\lfloor N_ns\rfloor}{N_n}G_n(1).
\]
We show that \(G_n(1)\) and \(G_n^{\mathrm{br}}\) approximate \(U_n\) and \(B_n\), respectively. 
\begin{theorem}[Gaussian approximation]
\label{thm:master-gaussian}
Suppose Assumptions~\ref{asm:data-score}, \ref{asm:covariance-nondegeneracy}, and \ref{asm:block-complexity}(i) hold. For every sequence of deterministic terminal transformations \(\mathsf A_n\) whose rows belong to \(\mathcal T_n\) and have uniformly bounded \(\ell_1\)-norms, we have
\[
\sup_{u\in\mathbb R}\left|\mathbb{P}\{\|\mathsf A_nU_n\|_\infty\le u\}-\mathbb{P}\{\|\mathsf A_nG_n(1)\|_\infty\le u\}\right|\to0.
\]
For the temporal bridge, we have
\[
\sup_{u\in\mathbb R}\left|\mathbb{P}\left\{\sup_{s\in\mathcal S_n(\epsilon)}\|B_n(s)\|_\infty\le u\right\}-\mathbb{P}\left\{\sup_{s\in\mathcal S_n(\epsilon)}\|G_n^{\mathrm{br}}(s)\|_\infty\le u\right\}\right|\to0.
\]
\end{theorem}

Write $\mathbb P^*$ for probability conditional on the observations. The bootstrap vectors and bridges are conditionally Gaussian, so their approximation reduces to covariance comparison. Lemma~\ref{lem:app-block-covariance} in the Supplementary Material bounds the covariance errors, yielding the following result under Assumption~\ref{asm:covariance-nondegeneracy}.

\begin{theorem}[Common block-bootstrap approximation]
\label{thm:master-bootstrap}
Suppose Assumptions~\ref{asm:data-score}, \ref{asm:covariance-nondegeneracy}, and \ref{asm:block-complexity}(ii) hold. For every sequence of deterministic terminal transformations \(\mathsf A_n\) whose rows belong to \(\mathcal T_n\) and have uniformly bounded \(\ell_1\)-norms, we have
\[
\sup_{u\in\mathbb R}\left|\mathbb{P}^*\{\|\mathsf A_nU_n^*\|_\infty\le u\}-\mathbb{P}\{\|\mathsf A_nG_n(1)\|_\infty\le u\}\right|\xrightarrow{p}0.
\]
For the temporal bridge, we have
\[
\sup_{u\in\mathbb R}\left|\mathbb{P}^*\left\{\sup_{s\in\mathcal S_n(\epsilon)}\|B_n^*(s)\|_\infty\le u\right\}-\mathbb{P}\left\{\sup_{s\in\mathcal S_n(\epsilon)}\|G_n^{\mathrm{br}}(s)\|_\infty\le u\right\}\right|\xrightarrow{p}0.
\]
\end{theorem}

Together, the two theorems provide the bootstrap approximations used for inference in Section~\ref{subsec:theory-inference}. For geometric regularity, replacing the population Jacobian by \(D\Phi_n(\widehat\Theta_n)\) additionally requires the linearization and plug-in error bounds established in Section~\ref{subsec:theory-ar}.

\subsection{Inferential guarantees}
\label{subsec:theory-inference}
We now establish size and power guarantees for the three tests in Section~\ref{subsec:method-structural}, together with guarantees for plateau cutoff selection and change-point localization. Throughout, $\alpha\in(0,1)$ is fixed, and the assumptions hold uniformly in $n$ along sequences of alternatives. Critical values are ideal conditional bootstrap quantiles; their Monte Carlo approximations have the same asymptotic interpretation when the number of replications tends to infinity.
Propositions~\ref{prop:stab-pointwise}, \ref{prop:cp-bootstrap}, and \ref{prop:geometric-bootstrap} establish asymptotic size $\alpha$, while Propositions~\ref{prop:stab-power} and \ref{prop:geometric-power} and Theorem~\ref{thm:cp-detection} establish consistency under the stated signal conditions. Proposition~\ref{prop:stab-cutoff} characterizes the first nonrejected stabilization cutoff, and Theorem~\ref{thm:cp-localization} gives a convergence rate for the estimated change-point location.


\subsubsection{Inference for profile stabilization}
\label{subsec:theory-stabilization}

Fix \(K_n\equiv K\ge2\) and a prespecified cutoff \(q_0\in\{0,\ldots,K-2\}\). Under \(H_0^{\mathrm{stab}}(q_0)\), Theorems~\ref{thm:master-gaussian} and \ref{thm:master-bootstrap}, applied with \(\mathsf A_n=C_{m_{q_0}}H_{m_{q_0}}P_{q_0}\), approximate the observed and bootstrap statistics by the same Gaussian maximum, yielding the following size guarantee.
\begin{proposition}[Bootstrap validity]
\label{prop:stab-pointwise}
Suppose Assumptions~\ref{asm:data-score}--\ref{asm:block-complexity} hold. Under \(H_0^{\mathrm{stab}}(q_0)\), we have
\[
  \sup_{u\in\mathbb R}\left|\mathbb{P}^*\{R_n^*(q_0)\le u\}-\mathbb{P}\{R_n(q_0)\le u\}\right|\xrightarrow{p}0.
\]
Consequently, \(\mathbb{P}\left\{R_n(q_0)>\widehat c_{n,1-\alpha}^{\mathrm{stab}}(q_0)\right\}\to\alpha.\)
\end{proposition}

Under an alternative, the reverse triangle inequality gives
\[
  R_n(q_0)\ge\sqrt{N_n}\|\mathsf A_n\Theta_n\|_\infty-\|\mathsf A_nU_n\|_\infty,
  \qquad \mathsf A_n=C_{m_{q_0}}H_{m_{q_0}}P_{q_0}.
\]
For fixed $K$ and $b_n=o(N_n)$, the moment bounds in Lemma~\ref{lem:app-fixed-window-moments} of the Supplementary Material imply that the stochastic term and bootstrap critical value are $O_p(1)$. A diverging scaled population contrast therefore yields the following.

\begin{proposition}[Consistency]
\label{prop:stab-power}
Suppose Assumption~\ref{asm:data-score} holds under \(H_a^{\mathrm{stab}}(q_0)\), with \(b_n=o(N_n)\). If \(\sqrt{N_n}\|C_{m_{q_0}}H_{m_{q_0}}P_{q_0}\Theta_n\|_\infty\to\infty\), then \(\mathbb{P}\{R_n(q_0)>\widehat c_{n,1-\alpha}^{\mathrm{stab}}(q_0)\}\to1\).
\end{proposition}

This includes every fixed nonconstant tail satisfying the assumptions, whose positive contrast norm diverges when scaled by \(\sqrt{N_n}\). 
The pointwise size and power results also characterize the cutoff estimator, provided that every candidate preceding the first plateau satisfies the signal condition.

\begin{proposition}[Stabilization cutoff selection]
\label{prop:stab-cutoff}
Suppose Assumptions~\ref{asm:data-score}--\ref{asm:block-complexity} hold uniformly in $n$ for all candidate cutoffs. Define $q_\star:=\min\{q_0\in\{0,\ldots,K-2\}:\theta_{q_0+1}=\cdots=\theta_K\}$, and assume that $q_\star$ does not depend on $n$. Suppose $\sqrt{N_n}\|C_{m_{q_0}}H_{m_{q_0}}P_{q_0}\Theta_n\|_\infty\to\infty$ for every candidate $q_0<q_\star$. For fixed $\alpha\in(0,1)$, if $q_\star<\infty$, then $\mathbb P(\widehat q_{n,\alpha}<q_\star)\to0$ and $\mathbb P(\widehat q_{n,\alpha}=q_\star)\to1-\alpha$. If $q_\star=\infty$, then $\mathbb P(\widehat q_{n,\alpha}=\infty)\to1$.
\end{proposition}

For a finite $q_\star$, the estimator asymptotically avoids premature truncation but overestimates the cutoff with probability tending to $\alpha$, reflecting the fixed-level rejection probability at the first true plateau.


\subsubsection{Inference for temporal stability}
\label{subsec:theory-cp}

Under the stationary models satisfying \(H_0^{\mathrm{temp}}\) and the stated assumptions, full-sample centering removes the common lag profile, and \(T_n(K_n,\epsilon)\) is exactly the bridge maximum covered by Section~\ref{subsec:theory-master}. Applying the bridge parts of Theorems~\ref{thm:master-gaussian} and \ref{thm:master-bootstrap} yields the following.

\begin{proposition}[Bootstrap validity]
\label{prop:cp-bootstrap}
Suppose Assumptions~\ref{asm:data-score}--\ref{asm:block-complexity} hold. Under \(H_0^{\mathrm{temp}}\), we have
\[
  \sup_{u\in\mathbb R}\left|\mathbb{P}^*\{T_n^*(K_n,\epsilon)\le u\}-\mathbb{P}\{T_n(K_n,\epsilon)\le u\}\right|\xrightarrow{p}0.
\]
Consequently, \(\mathbb{P}\{T_n(K_n,\epsilon)>\widehat c_{n,1-\alpha}^{\mathrm{CP}}\}\to\alpha.\)
\end{proposition}

To study detection and localization, consider a single change separating two stationary regimes at \(m_0=\lfloor N_n\tau_0\rfloor\), where \(\tau_0\in(\epsilon,1-\epsilon)\). Specifically, let \(X_t=X_t^{(1)}\) for \(t\le m_0\) and \(X_t=X_t^{(2)}\) for \(t>m_0\), where each \(\{X_t^{(r)}:t\in\mathbb Z\}\), \(r=1,2\), is stationary. Write their lag profiles as \(\Theta_n^{(1)}=(\theta_1^{(1)},\ldots,\theta_{K_n}^{(1)})^\top\) and \(\Theta_n^{(2)}=(\theta_1^{(2)},\ldots,\theta_{K_n}^{(2)})^\top\). Apart from discretization and lag pairs crossing the break, the mean of \(B_n(s)\) is \(\sqrt{N_n}\mathcal D_n(s)\), where
\begin{equation}
  \mathcal D_n(s):=
  \begin{cases}
    s(1-\tau_0)(\Theta_n^{(1)}-\Theta_n^{(2)}),&s\le\tau_0,\\
    \tau_0(1-s)(\Theta_n^{(1)}-\Theta_n^{(2)}),&s>\tau_0.
  \end{cases}
  \label{eq:cp-deterministic-drift}
\end{equation}
Let
\[
  \Delta_{n,h}:=\|\Theta_n^{(1)}-\Theta_n^{(2)}\|_\infty.
\]
The drift norm is maximized at \(\tau_0\), with height \(\tau_0(1-\tau_0)\Delta_{n,h}\). Under uniform moment bounds, the crossing-pair and discretization contributions to the mean bridge are \(O(K_n/\sqrt{N_n})\), uniformly over the scan. Let $p_n:=|\mathcal S_n(\epsilon)|K_n$ denote the number of lag--time coordinates in the temporal scan. The following theorem compares the drift at the change point with the sampling fluctuation and the bootstrap critical value under the alternative.

\begin{theorem}[Detection consistency]
\label{thm:cp-detection}
Suppose that both \(\{X_t^{(1)}:t\in\mathbb Z\}\) and \(\{X_t^{(2)}:t\in\mathbb Z\}\) satisfy Assumptions~\ref{asm:data-score} and \ref{asm:block-complexity}, with the long-run covariance and population block-covariance bias evaluated within each regime. Furthermore, \(q,\beta,\gamma,x_0\) and the tuning sequences \(K_n,b_n\) are common to both regimes, and all regularity bounds hold uniformly in \(n\) and across regimes. If
\[
\frac{\sqrt{N_n}\Delta_{n,h}}{\left(1+K_n/\sqrt{N_n}\right)\sqrt{\log p_n}}\to\infty,
\]
then \(\mathbb{P}\{T_n(K_n,\epsilon)>\widehat c_{n,1-\alpha}^{\mathrm{CP}}\}\to1\).
\end{theorem}
The signal condition requires the change in the lag profile to dominate both the sampling fluctuation and the contribution of pairs crossing the change point. The proof bounds the bootstrap critical value by $O_p\{(1+K_n/\sqrt{N_n})\sqrt{\log p_n}+\sqrt{b_n}\Delta_{n,h}\}$, which is negligible relative to the signal under the stated condition.

For localization, the relevant comparison is the drop in drift away from its maximizer. Define
\[
  \kappa_n(\eta):=\|\mathcal D_n(\tau_0)\|_\infty-\sup_{s\in\mathcal S_n(\epsilon):\,|s-\tau_0|\ge\eta}\|\mathcal D_n(s)\|_\infty,
\]
for \(\eta\) with a nonempty comparison set. The triangular drift in \eqref{eq:cp-deterministic-drift} gives a separation of order \(\eta\Delta_{n,h}\), up to grid effects. When this separation dominates the stochastic fluctuation and crossing-pair error, the maximizer of the empirical bridge concentrates near \(\tau_0\), yielding the consistency and localization rate in the following theorem.

\begin{theorem}[Change-point localization]
\label{thm:cp-localization}
Under the conditions of Theorem~\ref{thm:cp-detection}, let \(\eta_n\downarrow0\) satisfy \(N_n\eta_n\to\infty\). If
\begin{equation}
  \frac{\sqrt{N_n}\,\kappa_n(\eta_n)}{\sqrt{\log p_n}+K_n/\sqrt{N_n}}\to\infty,
  \label{eq:cp-localization-separation}
\end{equation}
then \(|\widehat\tau_n-\tau_0|=O_p(\eta_n)\).
\end{theorem}

For the triangular drift in \eqref{eq:cp-deterministic-drift}, condition \eqref{eq:cp-localization-separation} is satisfied whenever \(\sqrt{N_n}\eta_n\Delta_{n,h}/\{\sqrt{\log p_n}+K_n/\sqrt{N_n}\}\to\infty\). 


\subsubsection{Inference for geometric regularity}
\label{subsec:theory-ar}

Fix \(K_n\equiv K\ge3\). The geometric-regularity statistic evaluates a quadratic map of the empirical lag profile, so applying the terminal approximation requires linearization. Lemma~\ref{lem:curvature-taylor} in the Supplementary Material gives an expansion whose remainder is bounded by a constant multiple of \(\sqrt{N_n}\|\widehat\Theta_n-\Theta_n\|_\infty^2\). Such a quadratic remainder is negligible under \(\sqrt{N_n}\)-consistency, giving \(\sqrt{N_n}\Phi_n(\widehat\Theta_n)=D\Phi_n(\Theta_n)U_n+o_p(1)\) under \(H_0^{\mathrm{curv}}\). The lemma also controls the plug-in Jacobian error, so Theorems~\ref{thm:master-gaussian} and \ref{thm:master-bootstrap} yield the following.

\begin{proposition}[Bootstrap validity]
\label{prop:geometric-bootstrap}
Suppose Assumptions~\ref{asm:data-score}--\ref{asm:block-complexity} hold. Under \(H_0^{\mathrm{curv}}\), the following approximations hold:
\begin{enumerate}
\item[(i)] \(\sup_{u\in\mathbb R}|\mathbb{P}\{A_n(K_n)\le u\}-\mathbb{P}\{\|D\Phi_n(\Theta_n)G_n(1)\|_\infty\le u\}|\to0\).
\item[(ii)] \(\sup_{u\in\mathbb R}|\mathbb{P}^*\{A_n^*(K_n)\le u\}-\mathbb{P}\{A_n(K_n)\le u\}|\xrightarrow{p}0\).
\end{enumerate}
Consequently, \(\mathbb{P}\{A_n(K_n)>\widehat c_{n,1-\alpha}^{\mathrm{curv}}(K_n)\}\to\alpha\).
\end{proposition}

Under \(H_a^{\mathrm{curv}}\), the expansion becomes \(\sqrt{N_n}\Phi_n(\widehat\Theta_n)=\sqrt{N_n}\Phi_n(\Theta_n)+D\Phi_n(\Theta_n)U_n+o_p(1).\)
For fixed $K$ and $b_n=o(N_n)$, Lemmas~\ref{lem:app-fixed-window-moments} and \ref{lem:curvature-taylor} in the Supplementary Material imply that the linear fluctuation and bootstrap critical value are $O_p(1)$. Hence a diverging scaled population discrepancy yields consistency.

\begin{proposition}[Consistency]
\label{prop:geometric-power}
Suppose Assumption~\ref{asm:data-score} holds under \(H_a^{\mathrm{curv}}\) with \(b_n=o(N_n)\). If \(\sqrt{N_n}\|\Phi_n(\Theta_n)\|_\infty\to\infty\), then \(\mathbb{P}\{A_n(K_n)>\widehat c_{n,1-\alpha}^{\mathrm{curv}}(K_n)\}\to1\).
\end{proposition}
\section{Simulation Studies}
\label{sec:simulation}
We evaluate the proposed procedures for profile stabilization, temporal stability, and geometric regularity. All tests use level $\alpha=0.05$ and $2000$ Monte Carlo replications; the proposed procedures use $B=499$ block Gaussian multiplier draws with block length $b_n=\lceil 0.3K^{2/3}N_n^{1/3}\rceil$, where $N_n=n-K$ and $K$ is the lag horizon of each experiment. We take $n\in\{500,800,1200,1500\}$. Lower-tail QQ plots examine the null \(p\)-value distributions. Empirical rejection probabilities describe size under the null and power under the alternatives. 

\subsection{Profile stabilization}
\label{subsec:simulation-range}
We set the true stabilization lag to $q^\ast=5$, so that $q_0<q^\ast$ corresponds to the alternative and
$q_0\geq q^\ast$ to the null. For the Euclidean design, let $X_t\in\mathbb R^p$, with $p=100$, have independent Gaussian MA$(q^\ast)$ coordinates:
\[
X_{t,j}
=
\frac{1}{\sqrt{q^\ast+1}}
\sum_{\ell=0}^{q^\ast}\varepsilon_{t-\ell,j},
\qquad
\varepsilon_{t,j}\stackrel{\mathrm{i.i.d.}}{\sim}\N(0,1),
\quad j=1,\ldots,p.
\]
We use $h(x,y)=-\|x-y\|_2/\sqrt p$. Since $X_t$ and $X_{t+k}$ are independent for $k>q^\ast$, the lag profile stabilizes beyond $q^\ast$. We also apply the banded-dependence test of \citet{yao2018testing} to $W=X^\top$, treating the $p$ coordinates as independent replicates and the $n$ time points as ordered components; increasing $n$ thus increases the dimension of this test, while its sample size remains $p=100$. We set the bandwidth to $h_Y=q_0+1$, so that both tests share the same null--alternative boundary at $q_0=q^\ast$.

For the density-valued design, let $z_t=\sum_{\ell=0}^{q^\ast}w_\ell\varepsilon_{t-\ell}$, where $\varepsilon_t\stackrel{\mathrm{i.i.d.}}{\sim}\N(0,1)$ and the weights are proportional to $(1,0.4,\ldots,0.4,1)$ and normalized to have unit $\ell_2$ norm. We map $z_t$ to a two-component Gaussian mixture density via
\[
\mathcal G_c(z)(x)
=
\pi_c(z)\phi(x;-a,s^2)
+
\{1-\pi_c(z)\}\phi(x;a,s^2),
\qquad
\pi_c(z)=\{1+\exp(-cz)\}^{-1},
\]
where $\phi(\cdot;\mu,\sigma^2)$ denotes the Gaussian density with mean $\mu$ and variance $\sigma^2$; we set $a=1.5$ and $s=0.45$. We set $X_t=\mathcal G_1(z_t)$, evaluate the densities on 301 equally spaced grid points over $[-4,4]$, and renormalize them numerically on this grid. We use $h(f,g)=-d_H(f,g)$, where $d_H(f,g)=\{\frac12\int(\sqrt{f(x)}-\sqrt{g(x)})^2\,dx\}^{1/2}$. Since $z_t$ and $z_{t+k}$ are independent for $k>q^\ast$, so are $X_t$ and $X_{t+k}$, and hence the lag profile stabilizes beyond $q^\ast=5$.

\begin{figure}[t]
  \centering
  \includegraphics[width=0.75\textwidth]{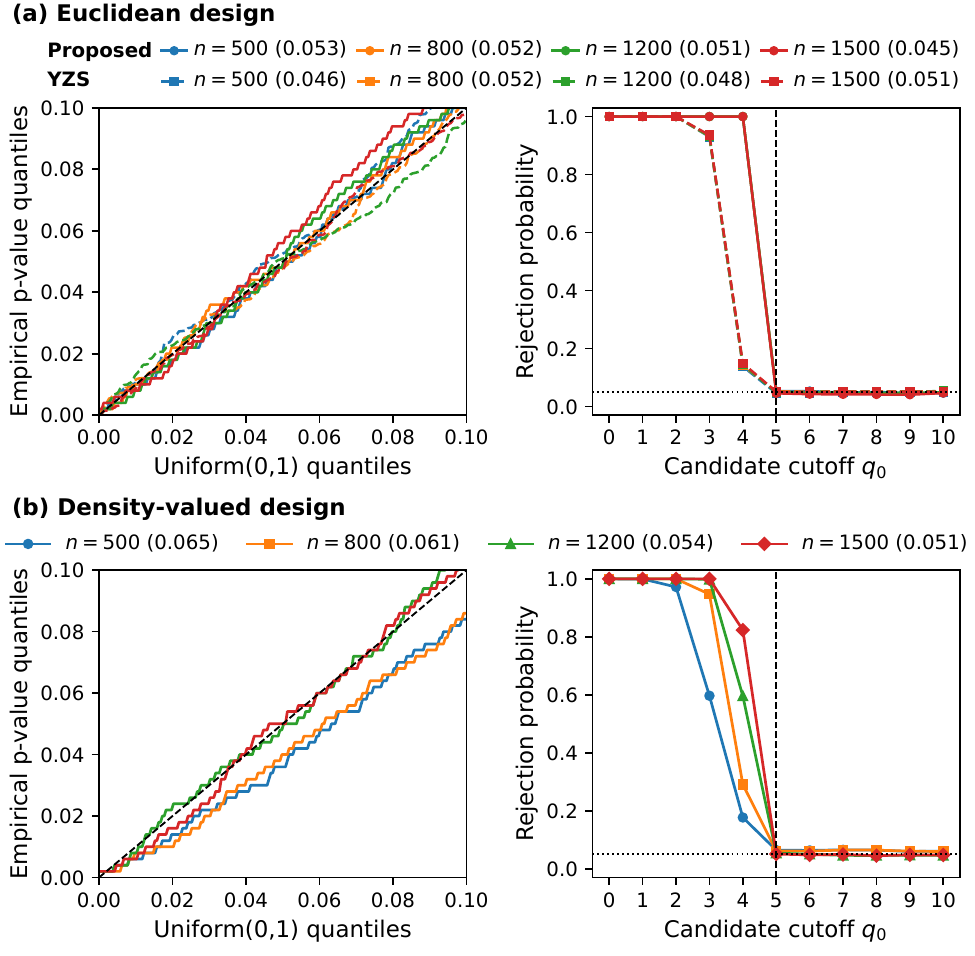}
  \caption{Profile-stabilization tests: (a) Euclidean design; (b) density-valued design. Left: null $p$-value QQ plots at $q_0=q^\ast$; right: rejection probabilities. Parentheses give empirical sizes at $\alpha=0.05$; reference lines mark the 45-degree line, $\alpha$, and $q^\ast$.}
  \label{fig:range-detection}
\end{figure}

For both designs, we set $K=20$ and test $q_0=0,\ldots,10$. In the Euclidean design (Figure~\ref{fig:range-detection}(a)), the lower-tail QQ curves of both tests lie close to the 45-degree line at every sample size, with empirical sizes between $0.045$ and $0.053$. Under the alternative, the proposed test maintains high power and attains higher rejection probabilities than the banded-dependence test as $q_0$ approaches $q^\ast$ from below. In the density-valued design (Figure~\ref{fig:range-detection}(b)), the QQ curves lie below the 45-degree line at $n=500$ and $800$, indicating mild over-rejection, and move close to it at $n=1200$ and $1500$; the power increases with $n$.

\subsection{Temporal stability}
\label{subsec:simulation-change-point}
We consider two complementary density-valued change-point designs with $m_0=\lfloor n/2\rfloor$: one changes the Hellinger Fr\'echet variance while preserving the Fr\'echet mean, whereas the other changes serial dependence while preserving the marginal distribution.

For the first design, let $\{\boldsymbol Z_t\}$ follow the stationary bivariate Gaussian AR(1) process $\boldsymbol Z_t=\rho\boldsymbol Z_{t-1} +\sqrt{1-\rho^2}\,\boldsymbol\varepsilon_t$, where $\boldsymbol\varepsilon_t\stackrel{\mathrm{i.i.d.}}{\sim}\N_2(\boldsymbol0,\boldsymbol I_2)$, $\boldsymbol Z_1\sim \N_2(\boldsymbol0,\boldsymbol I_2)$, and $\rho=0.2$. Set \(a_t=a_1\) for \(t\le m_0\) and \(a_t=a_2\) for \(t>m_0\), and define \(\boldsymbol\xi_t=a_t\boldsymbol Z_t/\sqrt{1+\|\boldsymbol Z_t\|_2^2}\). We construct $X_t=f_t$ on $[0,1]$ through the square-root density
\[
q_t(u)
=
\sqrt{1-\|\boldsymbol\xi_t\|_2^2}
+\sqrt{2}\,\xi_{t1}\cos(2\pi u)
+\sqrt{2}\,\xi_{t2}\sin(2\pi u),
\qquad
f_t(u)=q_t^2(u).
\]
Orthonormality ensures $\int_0^1 q_t^2(u)\,du=1$, and $\|\boldsymbol\xi_t\|_2<a_t<1/\sqrt{3}$ guarantees $q_t(u)>0$, so that $f_t=q_t^2$ is a valid density. By symmetry of $\boldsymbol Z_t$, the Hellinger Fr\'echet mean remains the uniform density for all $a_t$, while the corresponding Fr\'echet variance increases with $a_t$. We fix $a_1=0.15$ and take $a_2\in\{0.150,0.155,0.160,0.165,0.170,0.190,0.210\}$, with $\Delta_a=a_2-a_1$.

For the second design, let \(z_t=\rho_tz_{t-1}+\sqrt{1-\rho_t^2}\,\varepsilon_t\), where \(\varepsilon_t\stackrel{\mathrm{i.i.d.}}{\sim}\N(0,1)\), \(z_1\sim\N(0,1)\), and \(\rho_t=\rho_1\) for \(t\le m_0\) and \(\rho_t=\rho_2\) for \(t>m_0\). Then $z_t\sim \N(0,1)$ throughout, so changing $\rho_t$ alters serial dependence while preserving the marginal distribution. We set $X_t=\mathcal G_2(z_t)$ using the Gaussian-mixture map from Section~\ref{subsec:simulation-range}. We fix $\rho_1=0.2$ and take $\rho_2\in\{0.2,0.3,0.4,0.5,0.6,0.7\}$, with $\Delta_\rho=\rho_2-\rho_1$.

For both designs, we compare the proposed test with SS-SN \citep{zhang2026change} and SN1 and SN2 \citep{2024Jiangtwo}, using the Hellinger distance throughout. For the proposed test, we set $h(f,g)=-d_H(f,g)$, $K=10$, and trimming proportion $\epsilon=0.15$. For SS-SN, we use trimming fraction $0.15$ and the $5\%$ critical value $5.39$ tabulated in \citet{gao2025dimension}; for SN1 and SN2, we set $\eta_1=0.15$, $\eta_2=0.05$, and use the $5\%$ critical value $46.50$ reported in \citet{2024Jiangtwo}. QQ $p$-values for the competing methods use simulated pivotal null distributions, whereas size and power use these published critical values. 

\begin{figure}[t]
    \centering
    \includegraphics[width=0.75\textwidth]{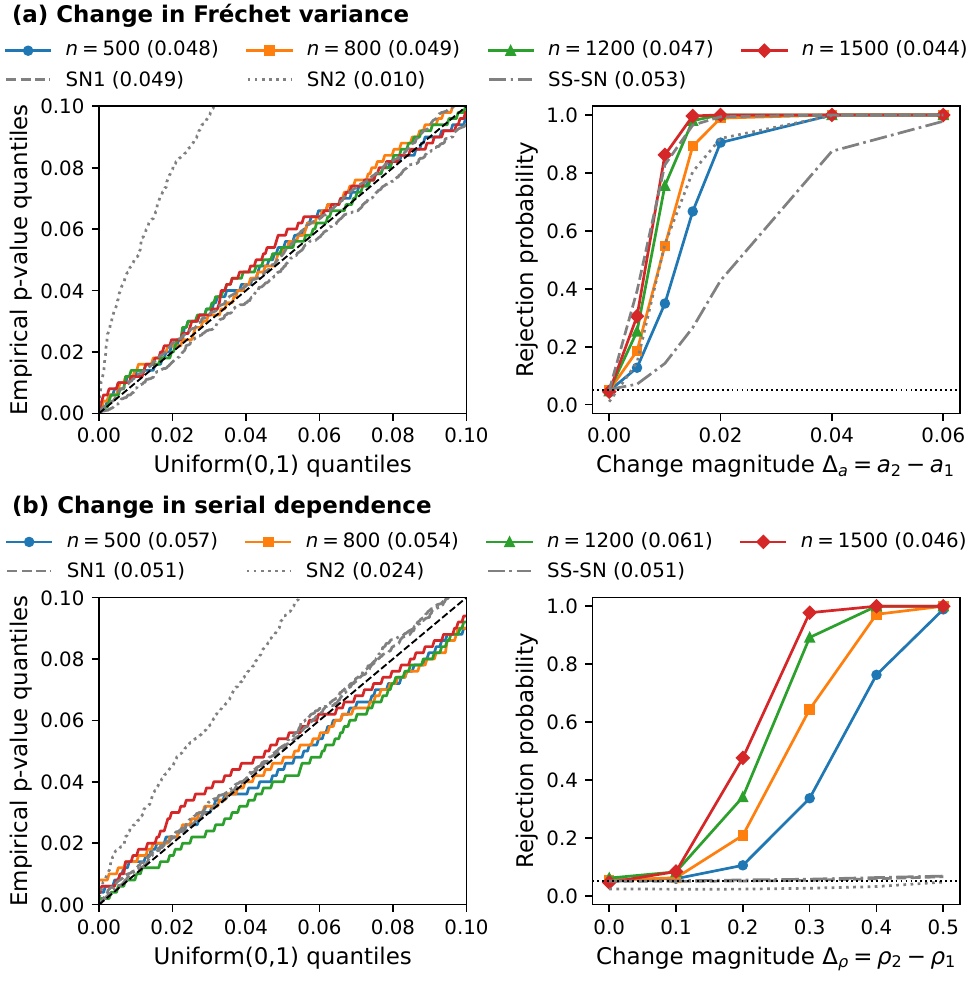}
    \caption{Temporal-stability tests: (a) change in Fr\'echet variance; (b) change in serial dependence. Left: null $p$-value QQ plots; right: rejection probabilities. Competing procedures (gray) are pooled across sample sizes.}
    \label{fig:change-point-density}
\end{figure}

In the Fr\'echet-variance design (Figure~\ref{fig:change-point-density}(a)), the QQ curves of the proposed test, SS-SN, and SN1 lie close to the 45-degree line, with empirical sizes of the proposed test between $0.044$ and $0.049$, while SN2 is conservative. Among the competitors, SN1, which targets this change directly, is the most powerful, and the proposed test is comparable to it for larger $n$. In the serial-dependence design (Figure~\ref{fig:change-point-density}(b)), the proposed test remains close to the nominal level, with empirical sizes between $0.046$ and $0.061$, and gains power with $\Delta_\rho$ and $n$, whereas the competing procedures remain near their null rejection levels, consistent with the unchanged marginal distribution.

\subsection{Geometric regularity}
\label{subsec:simulation-autoregression}
We consider Euclidean and density-valued designs with $c\in\{0,0.5,1,1.5,2,2.5,3\}$, where $c=0$ satisfies the geometric curvature restriction and $c>0$ controls the departure from it. For the Euclidean design, let $X_t\in\mathbb R^p$ with $p=20$ and $h(x,y)=p^{-1}x^\top y$. We generate the stationary Gaussian AR(2) process $X_t=0.2X_{t-1}+\phi_2X_{t-2}+\varepsilon_t$, where $\varepsilon_t\stackrel{\mathrm{i.i.d.}}{\sim}\N(0,I_p)$ and $\phi_2=0.035c$. At $c=0$, the process reduces to AR(1) with $\theta_k\propto0.2^k$.

For the density-valued design, partition $[0,1]$ into $2q$ equal-length bins, grouped into $q$ pairs $I_{r,+}$ and $I_{r,-}$, and define $g_r(x)=\sqrt q\{\mathbf 1(x\in I_{r,+})-\mathbf 1(x\in I_{r,-})\}$, $r=1,\ldots,q$. For each $r$, let $\{U_{r,t}\}$ and $\{V_{r,t}\}$ be stationary symmetric Markov chains on $\{-1,1\}$ satisfying $\E(U_{r,t}U_{r,t+k})=\rho_1^k$ and $\E(V_{r,t}V_{r,t+k})=\rho_2^k$, with all chains independent across $r$ and between families. We set $X_t=f_t$, where
\[
f_t(x)
=
1+\frac{\eta}{\sqrt q}\sum_{r=1}^q
\left\{U_{r,t}+\sqrt{\lambda(c)}V_{r,t}\right\}g_r(x),
\]
with $\lambda(c)=0.06c$. We take $q=200$, $\eta=0.5$, $\rho_1=0.2$, and $\rho_2=0.6$. Since each $g_r$ integrates to zero and $\lambda(c)\leq0.18$, $\int_0^1 f_t(x)\,dx=1$ and $f_t(x)\geq1-\eta\{1+\sqrt{\lambda(c)}\}>0$, so $f_t$ is a valid density. We use the centered $L^2$ inner product $h(f,g)=\int_0^1\{f(x)-1\}\{g(x)-1\}\,dx$. By orthonormality of
$\{g_r\}_{r=1}^q$ and independence of the two Markov-chain families, $\theta_k=\E\{h(f_t,f_{t+k})\}=\eta^2\rho_1^k+\eta^2\lambda(c)\rho_2^k$. Thus, $c=0$ yields a single geometric sequence, whereas $c>0$ introduces a second decay rate and violates the geometric curvature restriction.

\begin{figure}[t]
  \centering
  \includegraphics[width=0.75\textwidth]{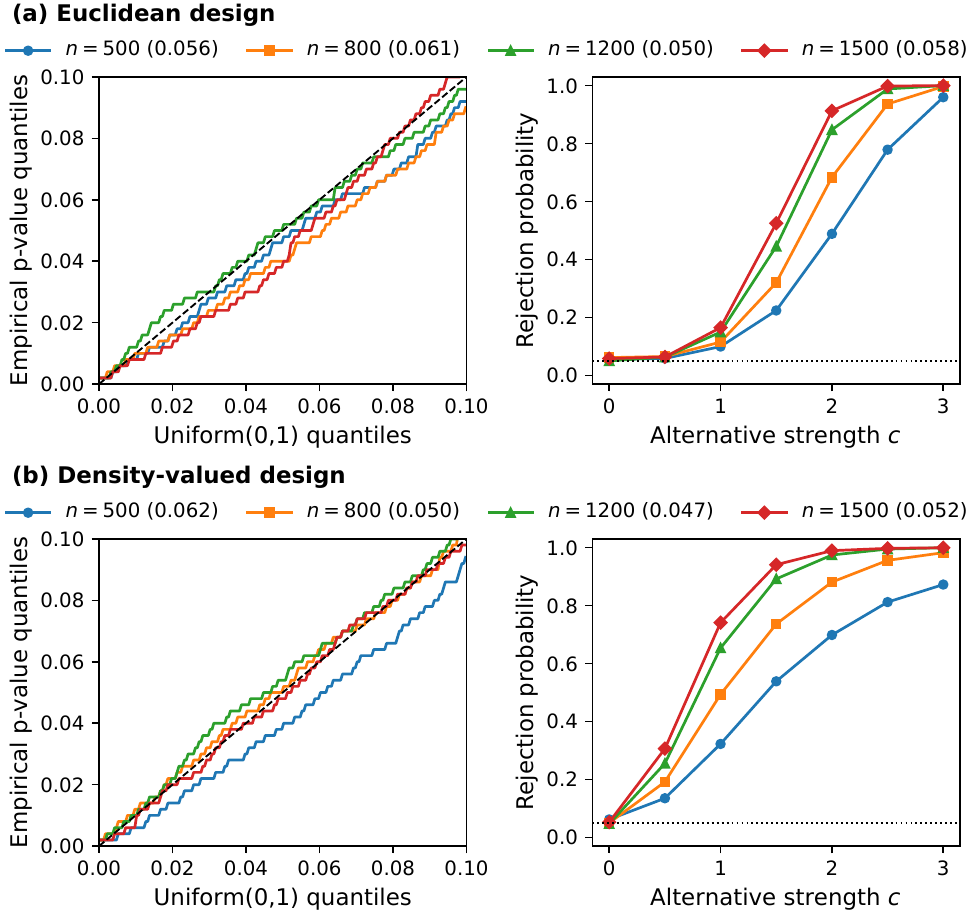}
  \caption{Geometric-regularity tests: (a) Euclidean design; (b) density-valued design. Left: null $p$-value QQ plots; right: rejection probabilities.}
  \label{fig:ar-detection}
\end{figure}

For both designs, we set $K=8$. Figure~\ref{fig:ar-detection} shows that the QQ curves lie close to the 45-degree line, with mild over-rejection in some cases: the empirical sizes range from $0.050$ to $0.061$ in the Euclidean design and from $0.047$ to $0.062$ in the density-valued design, the latter maximum occurring at $n=500$. The rejection probabilities increase in both $c$ and $n$ and approach one for larger values of $c$ and $n$.

\section{Real Data Applications}
\label{sec:real_data}
\subsection{Electricity load curves}
\label{subsec:electricity}
We analyze the UCI ElectricityLoadDiagrams20112014 dataset \citep{trindade2015electricity}, containing 15-minute electricity consumption measurements for 370 clients. Averaging consumption across clients and grouping by day yields daily load curves on a 96-point grid. We restrict the analysis to January 1, 2012--December 31, 2014, yielding $n=1096$ curves. After log transformation, spline smoothing, and removal of calendar effects (weekday, annual seasonality, and year), each day is represented by a smooth residual curve $X_t$ on $[0,24]$. Preprocessing details are given in Supplementary Section~\ref{supp:electricity-preprocessing}.

\begin{figure}[t]
    \centering
    \includegraphics[width=0.8\textwidth]{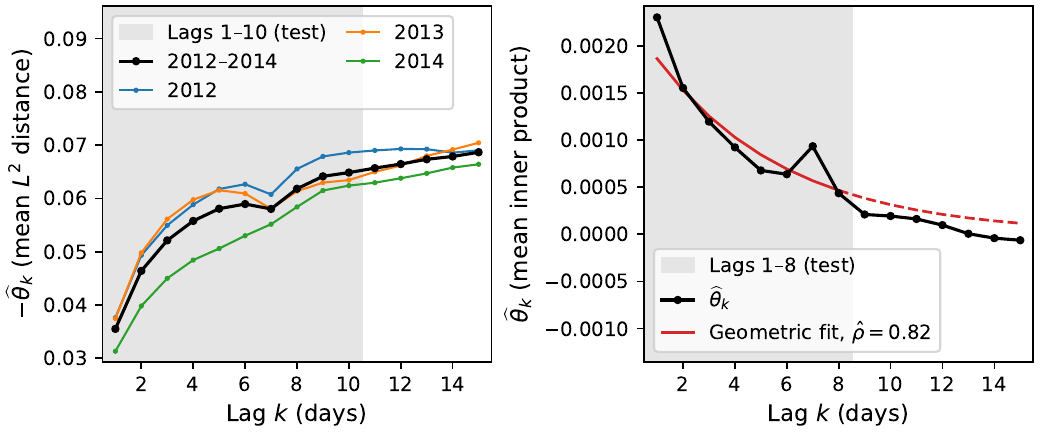}
    
    \caption{Empirical lag profiles for calendar-adjusted electricity load curves. Left: distance-based profile $-\widehat\theta_k$ for 2012--2014 and by calendar year of the anchor day. Right: inner-product profile $\widehat\theta_k$ with the geometric fit based on lags 1--8 (solid) and its extrapolation (dashed). In both panels, shading marks the lags used in the corresponding test.}
    \label{fig:electricity-lag-profile}
\end{figure}

We assess temporal stability and geometric regularity, using 4999 bootstrap replications and the block-length rule of Section~\ref{sec:simulation} for both procedures. For temporal stability, we use the distance-based pairwise function $h_{\mathrm{dist}}(x,y)=-\{24^{-1}\int_0^{24}[x(u)-y(u)]^2\,du\}^{1/2}$, so that $-\widehat\theta_k$ is the mean functional $L^2$ distance at lag $k$. The left panel of Figure~\ref{fig:electricity-lag-profile} shows that the mean distance generally increases with the lag and that the yearly profiles have similar shapes, with 2014 slightly lower in level. We set $K=10$ and $\epsilon=0.10$, giving $b_n=15$. The test does not reject temporal stability of the lag profile over the first 10 lags (statistic $0.1036$; 95\% critical value $0.1757$; $p=0.5504$). 

For geometric regularity, we use the functional inner product $h_{\mathrm{ip}}(x,y)=24^{-1}\int_0^{24}x(u)y(u)\,du$. Because the residual curves are centered, $\widehat\theta_k$ is an autocovariance-type summary at lag $k$. The right panel of Figure~\ref{fig:electricity-lag-profile} shows an inner-product profile that is close to the fitted geometric sequence with $\widehat\rho=0.82$ over the first 8 lags and falls to about zero beyond them. The local increase at lag 7 is consistent with weekly dependence. We set $K=8$, giving $b_n=13$. The test does not reject the geometric curvature restriction over the first 8 lags (statistic $1.9\times10^{-5}$; 95\% critical value $3.1\times10^{-5}$; $p=0.2200$).

Taken together, these two tests suggest a parsimonious description of the calendar-adjusted load curves. Temporal stability over the first 10 lags supports fitting a single dependence model to the full 2012--2014 period. Geometric regularity over the first 8 lags supports a geometric decay of the short-lag dependence. A first-order functional autoregression is therefore a natural candidate model.

\subsection{Cross-sectional stock-return distributions}
\label{subsec:stock-return-distributions}
We analyze cross-sectional return distributions for 78 large-cap U.S. stocks from January 2019 to December 2020, using adjusted closing prices from Yahoo Finance obtained via the \texttt{yfinance} Python package \citep{yfinance}. Daily log returns yield a balanced panel of $n=505$ trading days. For each trading day, we estimate the cross-sectional return density and treat it as the distribution-valued observation $X_t$. Density-estimation details are given in Supplementary Section~\ref{supp:stock-density}.

We assess temporal stability and, as an exploratory follow-up, profile stabilization, using 4999 bootstrap replications and the block-length rule of Section~\ref{sec:simulation} for both procedures. We use the negative Hellinger distance $h_{\mathrm H}(f,g)=-\{\frac12\int(\sqrt{f(x)}-\sqrt{g(x)})^2\,dx\}^{1/2}$, so that $-\widehat\theta_k$ is the mean Hellinger distance at lag $k$. For temporal stability, we set $K=10$ and $\epsilon=0.10$, giving $b_n=12$. The test rejects temporal stability (statistic $0.7246$; 95\% critical value $0.5699$; $p=0.0052$). The estimated break lies between February 24 and February 25, 2020, with the strongest contribution at lag $k=1$. For comparison, we also apply the SS-SN, SN1, and SN2 tests of Section~\ref{subsec:simulation-change-point} with the same tuning parameters and $5\%$ critical values. SS-SN rejects (statistic $7.65$; critical value $5.39$), with an estimated change point between February 5 and February 6, 2020, whereas SN1 and SN2 do not reject (statistics $10.33$ and $10.80$; critical value $46.50$). These procedures target changes in the marginal distribution and therefore capture a different feature of the series.

\begin{figure}[t]
    \centering
    \includegraphics[width=0.8\textwidth]{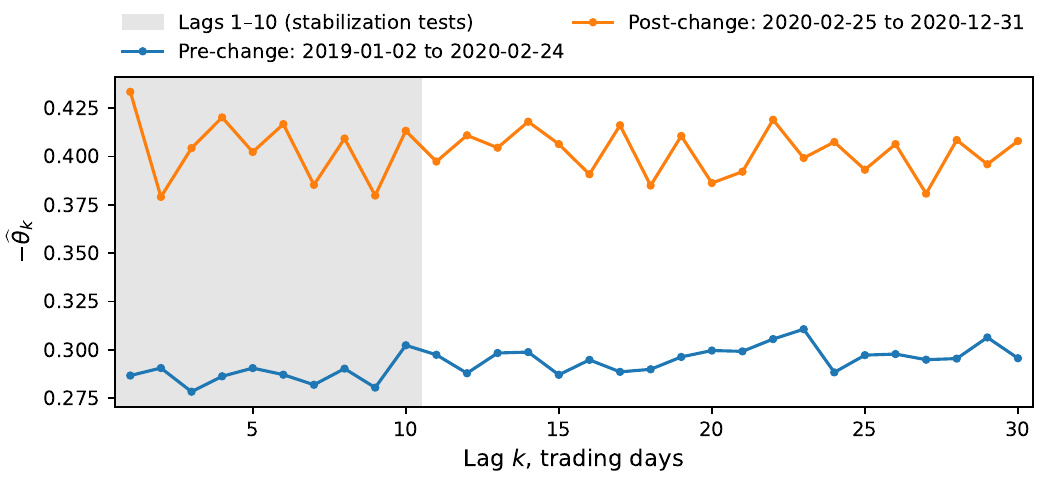}
    \caption{Pre- and post-change Hellinger lag profiles for the estimated cross-sectional stock-return densities. Shading marks the lags used in the profile-stabilization tests.}
    \label{fig:stock-pre-post}
\end{figure}

Using the break estimated by the temporal-stability test, we split the series into 288 pre-change trading days through February 24, 2020, and 217 post-change days beginning on February 25, 2020. Figure~\ref{fig:stock-pre-post} shows uniformly higher post-change mean Hellinger distances over lags $1$--$30$, indicating greater dissimilarity between cross-sectional return distributions after the break.

Based on the estimated split, we examine profile stabilization within each regime using $K=10$ and $q_0\in\{0,\ldots,8\}$. Applying the block-length rule to each regime gives $b_n=10$ and $9$ for the pre- and post-change segments. Table~\ref{tab:stock-range} reports pointwise $p$-values. In both regimes, constancy of the profile from lag 1 ($q_0=0$) is not rejected, and most cutoffs show no evidence against a constant tail; the few pointwise rejections occur at the largest cutoffs, where the tail contains only two to four lags.

\begin{table}[t]
\centering
\caption{Exploratory profile-stabilization tests for the pre- and post-change stock-return density series with $K=10$. Entries are pointwise $p$-values for the tail lags $q_0+1,\ldots,10$.}
\label{tab:stock-range}
\small
\setlength{\tabcolsep}{4pt}
\begin{tabular}{lccccccccc}
\toprule
$q_0$ & 0 & 1 & 2 & 3 & 4 & 5 & 6 & 7 & 8 \\
\midrule
Pre-change  & 0.5280 & 0.4170 & 0.3216 & 0.2884 & 0.1700 & 0.0530 & 0.0294 & 0.0354 & 0.0248 \\
Post-change & 0.1212 & 0.4216 & 0.0630 & 0.0716 & 0.2518 & 0.1482 & 0.2110 & 0.3180 & 0.0284 \\
\bottomrule
\end{tabular}
\end{table}

The estimated break between February 24 and February 25, 2020, coincides with the sharp movements in U.S. equity markets that accompanied the COVID-19 outbreak. \citet{baker2020unprecedented} identify news related to COVID-19 developments as the dominant driver of large daily U.S. stock market movements from February 24 through April 2020, providing a natural financial context for the detected change. Together with Figure~\ref{fig:stock-pre-post}, the exploratory stabilization results suggest that this change mainly reflects a higher level of day-to-day dissimilarity between cross-sectional return distributions rather than a change in how dissimilarity varies with temporal separation, in line with the frequent large market movements that followed the outbreak.

\section{Disclosure Statement}\label{disclosure-statement}
The authors have no conflict of interest to declare.

\section{Data Availability Statement}\label{data-availability-statement}

The code and processed data used in this work are available in an anonymous repository at \url{https://anonymous.4open.science/r/structural-analysis-object-valued-time-series-8B5D}. Instructions for obtaining the original data and reproducing the numerical results are provided in the repository.

\section{Declaration of Generative AI Use}
The authors used OpenAI Codex (model version: gpt-6-astra) and Anthropic Claude (model version: Opus 5.5) for language editing, assistance with mathematical proof checking, and code development and debugging, with the aim of improving clarity, identifying potential errors, and supporting computational implementation. The authors have reviewed the applicable terms of use and confirmed the suitability of the incorporated content for publication. The authors take full responsibility for the content of the manuscript.

\appendix


\clearpage
\forcsvlist{\renewSformat}{assumption,lemma,theorem,corollary,proposition,equation,section,table,figure,algocf}
\renewcommand*{\theHequation}{S.\arabic{equation}}

\begin{center}
{\large\bf Supplementary Material}
\end{center}

The supplementary material is organized as follows. Section~\ref{sec:supp-examples} presents examples of pairwise functions, lag-profile restrictions, and regularity conditions for pairwise functions. Section~\ref{sec:supp-numerical} gives the implementation algorithm, tuning guidance, a sensitivity analysis for the block length, and details of the real-data analyses. Section~\ref{sec:supp-results} collects notation, auxiliary lemmas and related results, and verification for the simulation constructions. Sections~\ref{sec:supp-theorems} and \ref{sec:supp-lemma-proofs} provide the proofs of the main theorems and propositions and of the auxiliary results, respectively.

\section{Examples of Pairwise Functions and Lag Profiles}
\label{sec:supp-examples}
\begin{table}[H]
\centering
\caption{Examples of pairwise functions and interpretations of the lag profile.}
\label{tab:pairwise-function-examples}
\begin{tabular}{p{0.25\textwidth} p{0.28\textwidth} p{0.37\textwidth}}
\toprule
Object/feature & Pairwise function \(h(x,y)\) & Interpretation of \(\theta_k\) \\
\midrule
Metric-space objects
&
\(d(x,y)\) or \(-d(x,y)\)
&
Average lag-\(k\) dissimilarity or similarity
\\[0.15cm]

Functional or Euclidean objects
&
\(\langle x,y\rangle\)
&
Average lag-\(k\) alignment
\\[0.15cm]

Kernel representations
&
\(\kappa(x,y)\), \quad \(\kappa\) positive definite
&
Kernel-based lag-\(k\) similarity
\\[0.15cm]

Application-specific objects
&
\(s(x,y)\in\mathbb R\)
&
Lag-\(k\) behavior of a problem-specific scalar feature
\\
\bottomrule
\end{tabular}
\end{table}

\begin{table}[p]
\centering
\begin{minipage}[t][0.94\textheight][s]{\textwidth}
\centering
\renewcommand{\baselinestretch}{1.2}\normalsize
\setlength{\tabcolsep}{6pt}
\renewcommand{\arraystretch}{1.25}
\captionsetup{font=small,skip=10pt}

\caption{Structural restrictions and the corresponding empirical inputs.}
\label{tab:structural-restrictions}
\begin{tabular}{@{}p{0.23\textwidth} p{0.34\textwidth} p{0.08\textwidth} p{\dimexpr0.35\textwidth-6\tabcolsep\relax}@{}}
\toprule
Task & Population restriction & Input & Transformation \\
\midrule
Profile stabilization & Tail profile is constant beyond $q_0$ & $U_n$ & Tail centering and cumulative contrast \\[0.15cm]
Temporal stability & Selected lag profile is constant over time & $B_n(s)$ & Lag--time sup norm \\[0.15cm]
Geometric regularity & Short-lag profile satisfies the geometric curvature restriction & $U_n$ & Plug-in Jacobian linearization \\
\bottomrule
\end{tabular}

\par\vspace{2cm}

\caption{Regularity of pairwise functions under Assumption~\ref{asm:data-score}. The indicated observation moment and physical-dependence order yield score moment order \(q\ge4\).}
\label{tab:score-regularity}
\begin{tabular}{@{}p{0.25\textwidth} p{0.23\textwidth} p{\dimexpr0.52\textwidth-4\tabcolsep\relax}@{}}
\toprule
Pairwise function and metric & Growth and moments & Examples and justification \\
\midrule
\(h(x,y)=\pm d(x,y)\) & \(\gamma=0\); order \(q\) & Triangle inequality; Euclidean and functional distances, Hellinger distance on densities. \\[0.2cm]
\(h(x,y)=\langle x,y\rangle\), norm metric & \(\gamma=1\); order \(2q\) & Norm-weighted Lipschitz bound; Gaussian AR and functional inner-product examples. \\[0.2cm]
\(h(f,g)=\langle f-1,g-1\rangle_{L^2}\) on a bounded set & \(\gamma=0\); order \(q\) & The norms \(\|f-1\|_2\) are uniformly bounded in the density simulation, giving a uniform Lipschitz constant. \\
\bottomrule
\end{tabular}

\end{minipage}
\end{table}

\newpage
\section{Auxiliary Numerical Results}
\label{sec:supp-numerical}

\subsection{Algorithm}
\begin{algorithm}[!htbp]
\caption{Unified implementation of lag-profile inference}
\label{alg:lag-profile-implementation}

\KwIn{Observations \(X_1,\ldots,X_n\); pairwise function \(h\); lag horizon \(K_n\); block length \(b_n\); bootstrap replications \(B\); significance level \(\alpha\); hypotheses \(\mathfrak S\), with cutoffs \(q_0\) and trimming parameter \(\epsilon\) as needed.}
\KwOut{Empirical lag profile, test statistics, bootstrap critical values, approximate \(p\)-values, and test decisions.}

Set \(N_n\gets n-K_n\) and \(L_n\gets\lceil N_n/b_n\rceil\)\;
Compute \(H_{t,n}\) and \(\widehat\Theta_n\) from \eqref{eq:lag-score-vector}--\eqref{eq:common-denom-estimator}, and \(B_n(s)\) from \eqref{eq:observable-bridge} if temporal stability is tested\;
Compute the selected statistics \(R_n(q_0)\), \(T_n(K_n,\epsilon)\), and/or \(A_n(K_n)\)\;
Form the blocks \(I_{\ell,n}\) in \eqref{eq:block-partition}\;

\For{\(b=1,\ldots,B\)}{
  Draw independent \(\N(0,1)\) multipliers \(\xi_1^{(b)},\ldots,\xi_{L_n}^{(b)}\), independently of the data and other replications\;
  Compute \(U_n^{*(b)}\) and/or \(B_n^{*(b)}(s)\) from \eqref{eq:bootstrap-partial-sum}--\eqref{eq:bootstrap-partial-sum-bridge}, using the same multipliers across all selected tests and scan points\;
  Compute the corresponding bootstrap statistics as defined in Section~\ref{subsec:method-structural}\;
}

\ForEach{\(\mathcal H\in\mathfrak S\)}{
  Compute the empirical \((1-\alpha)\)-quantile of its bootstrap statistics and reject \(\mathcal H\) if the observed statistic exceeds it\;
  Estimate its \(p\)-value by the proportion of bootstrap statistics at least as large as the observed statistic\;
}
\end{algorithm}

\subsection{Tuning and interpretation}\label{subsec:method-tuning}

The pairwise function $h$ and lag horizon $K_n$ determine the inferential target, whereas $b_n$ controls bootstrap calibration. The choice of $h$ should follow the scientific question, and $K_n$ should balance the lag range of interest against the effective sample size $N_n=n-K_n$.

A practical block-length choice is $b_n=\lceil cK_n^{2/3}N_n^{1/3}\rceil$ for a moderate constant $c>0$, as used in the numerical studies. Under Assumption~\ref{asm:data-score}, this rule satisfies Assumption~\ref{asm:block-complexity}(ii) for fixed \(K\) and, when \(K_n\asymp N_n^\kappa\), for \(0\le\kappa<q/(q+12)\). Section~\ref{supp:block-sensitivity} reports results under alternative block-length rules.

The trimming constant $\epsilon$ excludes candidate change points near the sample boundaries and should be fixed in advance. Searches over multiple cutoffs, lag horizons, or pairwise functions require multiplicity adjustment or an explicitly exploratory interpretation.

\subsection{Sensitivity to the block length}
\label{supp:block-sensitivity}
The simulations in Section~\ref{sec:simulation} use $b_n=\lceil 0.3K^{2/3}N_n^{1/3}\rceil$. To assess sensitivity to this choice, and in particular to block lengths that do not exceed the lag horizon, we repeat all six experiments with the alternative rules in Table~\ref{tab:block-sensitivity}, keeping all other settings, including the random seeds, unchanged. The alternative rules give $b_n<K$ in the profile-stabilization experiments at every sample size, $b_n=K$ in the geometric-regularity experiments, and slightly longer block lengths than the main rule in the temporal-stability experiments.

\begin{table}[htbp]
\centering
\caption{Lag horizons and block lengths under the main rule $\lceil 0.3K^{2/3}N_n^{1/3}\rceil$ and the alternative rules.}
\label{tab:block-sensitivity}
\footnotesize
\begin{tabular}{lclcccc}
\toprule
Experiment & $K$ & Rule & $n=500$ & $n=800$ & $n=1200$ & $n=1500$ \\
\midrule
Profile stabilization & 20 & Main & 18 & 21 & 24 & 26 \\
                      &    & Alternative: $\lceil1.6N_n^{1/3}\rceil$ & 13 & 15 & 17 & 19 \\
\midrule
Temporal stability    & 10 & Main & 11 & 13 & 15 & 16 \\
                      &    & Alternative: $\lceil1.5N_n^{1/3}\rceil$ & 12 & 14 & 16 & 18 \\
\midrule
Geometric regularity  & 8  & Main & 10 & 12 & 13 & 14 \\
                      &    & Alternative: $\max\{8,\lceil1.2N_n^{1/4}\rceil\}$ & 8 & 8 & 8 & 8 \\
\bottomrule
\end{tabular}
\end{table}

\begin{figure}[htbp]
  \centering
  \includegraphics[width=0.65\textwidth]{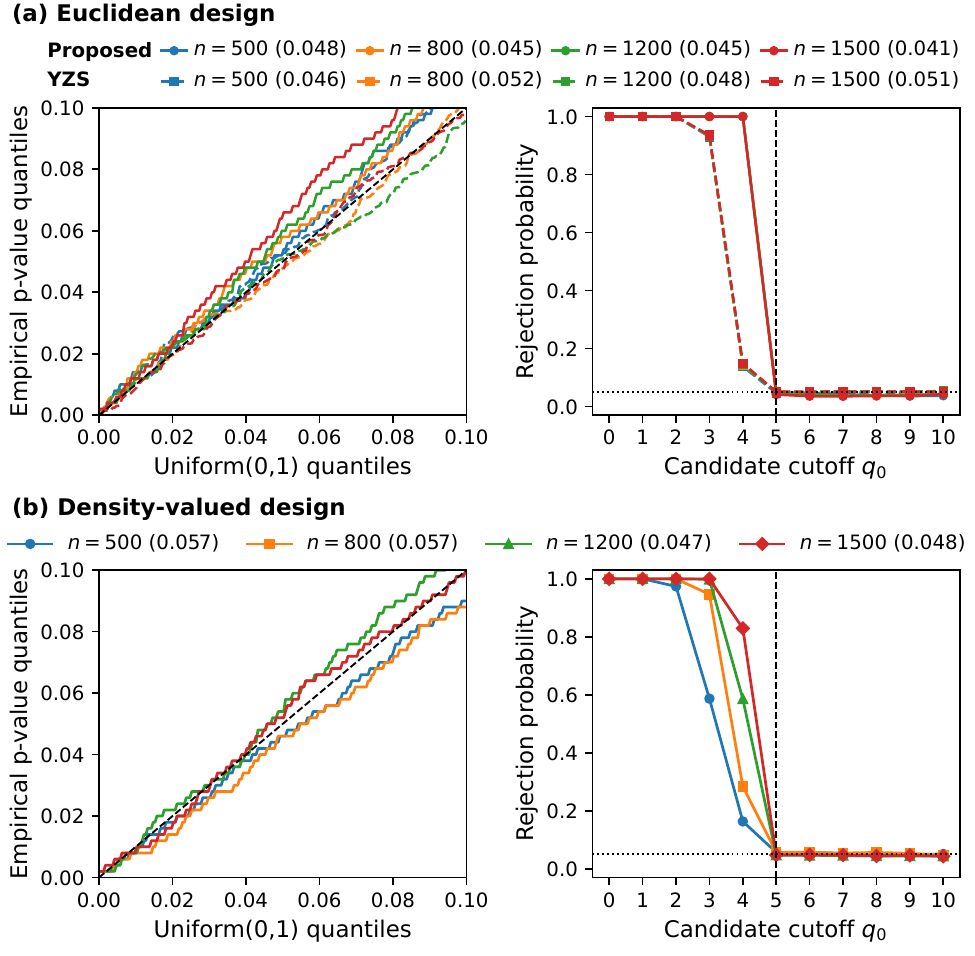}
  \caption{Profile-stabilization tests as in Figure~\ref{fig:range-detection}, with $b_n=\lceil1.6N_n^{1/3}\rceil$.}
  \label{fig:sens-range}
\end{figure}

\begin{figure}[htbp]
  \centering
  \includegraphics[width=0.65\textwidth]{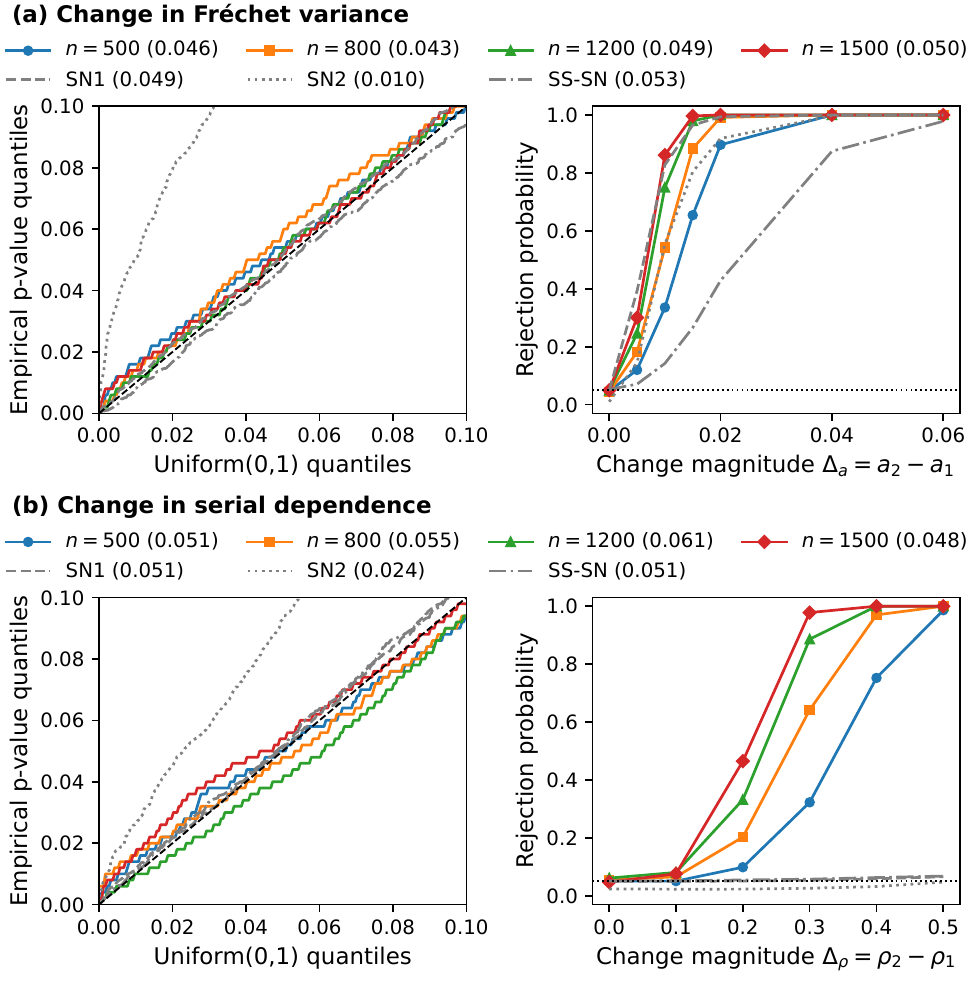}
  \caption{Temporal-stability tests as in Figure~\ref{fig:change-point-density}, with $b_n=\lceil1.5N_n^{1/3}\rceil$.}
  \label{fig:sens-tem}
\end{figure}

\begin{figure}[htbp]
  \centering
  \includegraphics[width=0.65\textwidth]{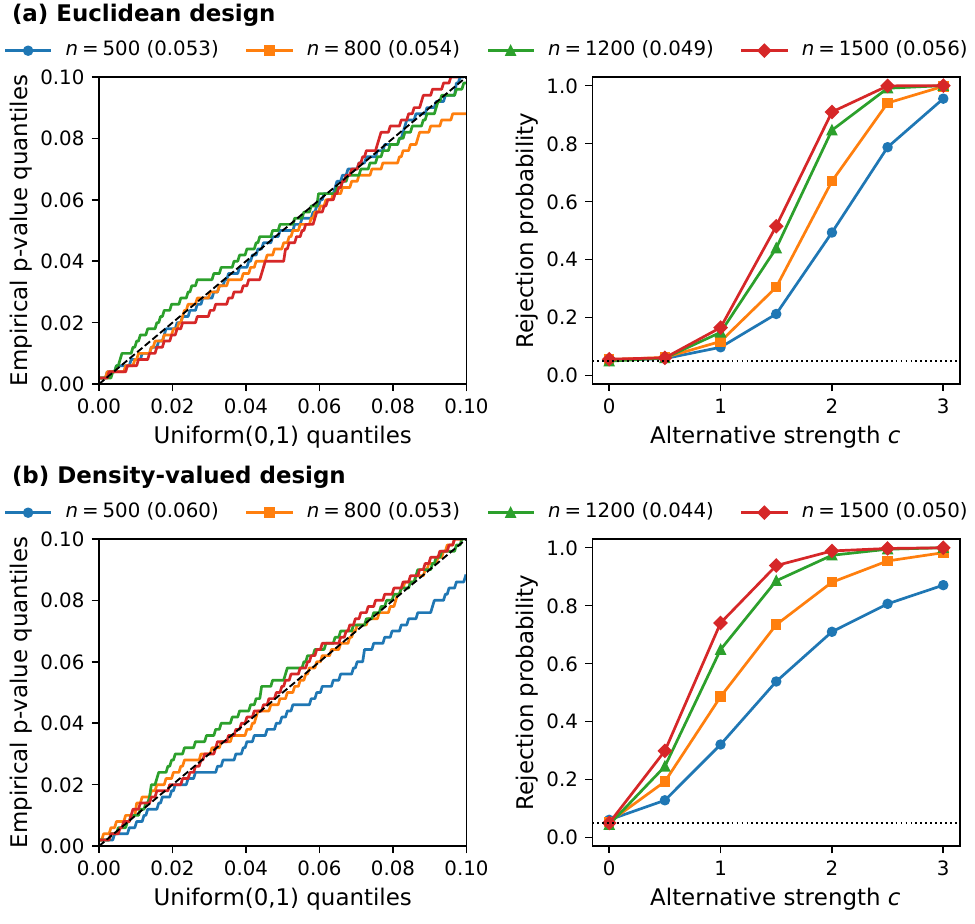}
  \caption{Geometric-regularity tests as in Figure~\ref{fig:ar-detection}, with $b_n=\max\{8,\lceil1.2N_n^{1/4}\rceil\}$.}
  \label{fig:sens-ar}
\end{figure}

Figures~\ref{fig:sens-range}--\ref{fig:sens-ar} report the results. In the profile-stabilization experiments, $K/b_n$ ranges from $1.05$ to $1.54$. The empirical sizes of the proposed test at the boundary null $q_0=q^\ast$ are between $0.041$ and $0.048$ in the Euclidean design, with the test slightly conservative at $n=1500$, and between $0.047$ and $0.057$ in the density-valued design. The power curves are close to those in Figure~\ref{fig:range-detection}. In the temporal-stability and geometric-regularity experiments, the empirical sizes of the proposed tests are between $0.043$ and $0.061$ and between $0.044$ and $0.060$, respectively, and the power curves are close to those in Figures~\ref{fig:change-point-density} and~\ref{fig:ar-detection}. Therefore, the finite-sample size and power of the three tests are similar under the two block-length rules. In particular, block lengths shorter than the lag horizon do not inflate the size of the profile-stabilization test relative to the main rule.

\subsection{Preprocessing details for real data}
\label{sec:supp-applications}

This section describes the smoothing, calendar adjustment, and density estimation used to construct the analyzed object sequences.

\subsubsection{Electricity load curves}
\label{supp:electricity-preprocessing}
For day $t$ and $j=1,\ldots,96$, let $Y_t(u_j)$ denote the 15-minute electricity load at time $u_j=(j-1)/4$ hours, averaged across the 370 clients, and let $\widetilde Y_t(u_j)=\log\{1+Y_t(u_j)\}$. We represent each daily curve in a common cubic B-spline basis $B_1,\ldots,B_M$ on $[0,24]$ with $M=20$ and equally spaced interior knots, and write $\boldsymbol B(u)=(B_1(u),\ldots,B_M(u))^\top$.

Let
\[
\boldsymbol R=\int_0^{24}\boldsymbol B''(u)\boldsymbol B''(u)^\top\,du
\]
denote the second-derivative roughness matrix, evaluated numerically by trapezoidal quadrature on 2001 equally spaced points over $[0,24]$. Let $\mathcal J_t\subseteq\{1,\ldots,96\}$ index the observations retained for day $t$: $\mathcal J_t=\{1,\ldots,96\}$, except on the six daylight-saving transition dates between 2012 and 2014, where the four observations from 01:00 to 01:45 are omitted. Let $\boldsymbol B$ be the $96\times M$ matrix with rows $\boldsymbol B(u_j)^\top$, let $\boldsymbol B_t$ be the submatrix of $\boldsymbol B$ formed by the rows $j\in\mathcal J_t$, and let $\widetilde{\boldsymbol Y}_t=(\widetilde Y_t(u_j))_{j\in\mathcal J_t}$ be the corresponding vector of observations. For a fixed smoothing parameter $\lambda$, the coefficient vector is estimated by
\[
\widehat{\boldsymbol c}_t = \arg\min_{\boldsymbol c} \left\{ \|\widetilde{\boldsymbol Y}_t-\boldsymbol B_t\boldsymbol c\|_2^2 + \lambda\boldsymbol c^\top \boldsymbol R\boldsymbol c \right\}.
\]

A common smoothing parameter is selected by pooled generalized cross-validation using the complete 96-point transformed curves. Let $\widetilde{\boldsymbol Y}_t^{\circ}=(\widetilde Y_t(u_1),\ldots,\widetilde Y_t(u_{96}))^\top$ and $\boldsymbol S_\lambda=\boldsymbol B(\boldsymbol B^\top\boldsymbol B+\lambda \boldsymbol R)^{-1}\boldsymbol B^\top$. The criterion is
\[
\operatorname{GCV}(\lambda)
=
\frac{\overline{\operatorname{RSS}}(\lambda)/96}{\{1-\operatorname{tr}(\boldsymbol S_\lambda)/96\}^2},
\qquad
\overline{\operatorname{RSS}}(\lambda)
=
\frac1n\sum_{t=1}^{n}\big\|\widetilde{\boldsymbol Y}_t^{\circ}-\boldsymbol S_\lambda\widetilde{\boldsymbol Y}_t^{\circ}\big\|_2^2,
\]
where $\operatorname{tr}(\boldsymbol S_\lambda)$ is the effective degrees of freedom. We minimize this criterion over 121 logarithmically spaced candidate values obtained by multiplying $\operatorname{tr}(\boldsymbol B^\top\boldsymbol B)/\operatorname{tr}(\boldsymbol R)$ by values ranging from $10^{-6}$ to $10^6$. The selected value is used in the fits for all days, including the daylight-saving transition dates. The resulting smooth curve is
\[
\widehat Y_t(u)=\boldsymbol B(u)^\top\widehat{\boldsymbol c}_t=\sum_{m=1}^{M}\widehat c_{t,m}B_m(u),
\qquad u\in[0,24].
\]

The smoothed curves $\widehat Y_t$ exhibit calendar effects, including weekly and annual seasonal variation and differences in the overall level between years. Before applying the proposed tests, we remove these effects. Let $d(t)\in\{1,\ldots,7\}$, $\tau_t$, and $y(t)$ denote the weekday, the day of the year, and the calendar year of day $t$. We collect the calendar covariates in $\boldsymbol z_t=\big(\boldsymbol w_t^\top,\boldsymbol s_t^\top,\boldsymbol v_t^\top\big)^\top\in\mathbb R^{15}$, where
\begin{align*}
\boldsymbol w_t&=\big(\mathbf 1\{d(t)=1\},\ldots,\mathbf 1\{d(t)=7\}\big)^\top,\\
\boldsymbol s_t&=\big(\sin\omega_1\tau_t,\cos\omega_1\tau_t,\ldots,\sin\omega_3\tau_t,\cos\omega_3\tau_t\big)^\top,
\qquad \omega_\ell=2\pi\ell/365.25,\\
\boldsymbol v_t&=\big(\mathbf 1\{y(t)=2013\},\mathbf 1\{y(t)=2014\}\big)^\top
\end{align*}
contain the weekday indicators, three annual harmonics, and year indicators with 2012 as the reference year.

The calendar effects are estimated by least squares in coefficient space,
\[
\widehat{\boldsymbol\Gamma}
=
\arg\min_{\boldsymbol\Gamma\in\mathbb R^{M\times 15}}
\sum_{t=1}^{n}\big\|\widehat{\boldsymbol c}_t-\boldsymbol\Gamma\boldsymbol z_t\big\|_2^2,
\]
and the final functional time series used in the analysis consists of the residual curves
\[
X_t(u)=\boldsymbol B(u)^\top\widehat{\boldsymbol c}_t^{\,X},
\qquad
\widehat{\boldsymbol c}_t^{\,X}=\widehat{\boldsymbol c}_t-\widehat{\boldsymbol\Gamma}\boldsymbol z_t,
\qquad u\in[0,24].
\]

Let
\[
\boldsymbol G = \frac{1}{24} \int_0^{24} \boldsymbol B(u)\boldsymbol B(u)^\top\,du
\]
be the normalized B-spline Gram matrix. The functional $L^2$ distance is then evaluated as
\[
\left\{ \frac{1}{24} \int_0^{24} [X_t(u)-X_s(u)]^2\,du \right\}^{1/2} = \left\{\big(\widehat{\boldsymbol c}_t^{\,X}-\widehat{\boldsymbol c}_s^{\,X}\big)^{\top} \boldsymbol G\,\big(\widehat{\boldsymbol c}_t^{\,X}-\widehat{\boldsymbol c}_s^{\,X}\big)\right\}^{1/2}.
\]
Similarly, the functional inner product is evaluated as
\[
\frac{1}{24}\int_0^{24}X_t(u)X_s(u)\,du=(\widehat{\boldsymbol c}_t^{\,X})^\top \boldsymbol G\,\widehat{\boldsymbol c}_s^{\,X}.
\]

\subsubsection{Cross-sectional stock-return distributions}
\label{supp:stock-density}
We use adjusted closing prices for a fixed universe of 78 large-cap U.S. stocks from January 2019 to December 2020, obtained from Yahoo Finance through the \texttt{yfinance} Python package \citep{yfinance}. Let $P_{i,t}$ denote the adjusted closing price of stock $i$ on trading day $t$. The daily log return is defined as
\[
r_{i,t} = \log P_{i,t}-\log P_{i,t-1}.
\]
Restricting to trading days with complete observations for all 78 stocks yields a balanced panel of $n=505$ daily cross sections.

For each trading day $t$, we estimate the cross-sectional return density using a Gaussian kernel density estimator,
\[
\widehat f_t(x) = \frac{1}{78h_{\mathrm{KDE}}} \sum_{i=1}^{78} \varphi\!\left( \frac{x-r_{i,t}}{h_{\mathrm{KDE}}} \right),
\]
where $\varphi$ is the standard Gaussian density. A common bandwidth is used for all trading days to ensure that the estimated densities are smoothed on the same scale. We select the bandwidth by Silverman’s robust rule of thumb \citep[eq.~(3.31)]{silverman1986density}, $h_{\mathrm{KDE}}=0.9\min\{\hat\sigma,\widehat{\mathrm{IQR}}/1.34\}\,78^{-1/5}$, where $\hat\sigma$ and $\widehat{\mathrm{IQR}}$ are the standard deviation and interquartile range of the pooled returns from all stocks and trading days. This gives $h_{\mathrm{KDE}}=0.0052$. Each $\widehat f_t$ is evaluated on an equally spaced grid of 801 points on $[\min_{i,t} r_{i,t}-4h_{\mathrm{KDE}},\,\max_{i,t} r_{i,t}+4h_{\mathrm{KDE}}]$ and rescaled to integrate to one under the trapezoidal rule. The resulting distribution-valued observation used in the analysis is
\[
X_t=\widehat f_t, \qquad t=1,\ldots,505.
\]

\section{Auxiliary Theoretical Results}
\label{sec:supp-results}
\subsection{Notation}
\label{sec:supp-notation}
We retain the notation of the main text. For a random variable \(Y\) and \(u\ge1\), write \(\|Y\|_u:=(\E|Y|^u)^{1/u}\). The exponent \(q\ge4\) in Assumption~\ref{asm:data-score} refers to score moments, while the observation moment is of order \(q(1+\gamma)\). For a vector \(x\), \(\|x\|_\infty\) and \(\|x\|_2\) denote the maximum and Euclidean norms. For a matrix \(A\), \(\|A\|_{\max}\) denotes the largest absolute entry and \(\|A\|_{\infty\to\infty}:=\max_i\sum_j|A_{ij}|\). The symbols \(\mathbb{P}^*\), \(\E^*\), and \(\Cov^*\) denote probability, expectation, and covariance conditional on the observations. We write \(e_k\) for the \(k\)th canonical basis vector, \(\mathbf 1_m\) for the vector of \(m\) ones, \(\1\{\cdot\}\) for the indicator function, \(|A|\) for the cardinality of a finite set \(A\), \(a\vee b:=\max(a,b)\), \(a\wedge b:=\min(a,b)\), and \((x)_+:=\max\{x,0\}\). For nonnegative sequences, \(a_n\lesssim b_n\) means \(a_n\le Cb_n\) for all sufficiently large \(n\), and \(a_n\ll b_n\) means \(a_n=o(b_n)\). Convergence in probability is denoted by \(\xrightarrow{p}\), and \(O_p(\cdot)\) and \(o_p(\cdot)\) are the usual stochastic orders. For a bootstrap random variable \(Y_n^*\), the notation \(Y_n^*=o_p^*(1)\) in probability means \(\mathbb{P}^*(|Y_n^*|>\eta)\xrightarrow{p}0\) for every \(\eta>0\); the notation \(Y_n^*=O_p^*(1)\) in probability means that, for every \(\eta>0\), \(\lim_{M\to\infty}\limsup_{n\to\infty}\mathbb{P}\{\mathbb{P}^*(|Y_n^*|>M)>\eta\}=0\). The deterministic terminal transformation is denoted by \(\mathsf A_n\), whereas \(A_n(K_n)\) denotes the geometric-regularity statistic. Constants denoted by \(C\) may change between occurrences and are independent of \(n\), the lag coordinate, and the scan point, with uniformity across alternatives as specified in the main text.

\subsection{Auxiliary lemmas}
\label{sec:supp-auxiliary}
For the moment and covariance bounds, define $\widetilde W_{t,n}=(\widetilde W_{t,n,1},\ldots,\widetilde W_{t,n,K_n})^\top$, where $\widetilde W_{t,n,k}:=h(X_{t-k},X_t)-\theta_k$ for $1\le k\le K_n$. For $t\in\mathbb Z$ and $j\ge0$, let $X_s^{*(t,j)}$ denote the coupled version of $X_s$ obtained by replacing $\varepsilon_{t-j}$ by an independent copy, and define the coupled lag-score vector $\widetilde W_{t,n}^{*(j)}$ by $\widetilde W_{t,n,k}^{*(j)} := h\bigl(X_{t-k}^{*(t,j)},X_t^{*(t,j)}\bigr)-\theta_k$ for $  1\le k\le K_n$.

Let $\mathcal F_j:=\sigma(\ldots,\varepsilon_{j-1},\varepsilon_j)$ for $j\in\mathbb Z$, let $\mathcal F_{-\infty}:=\bigcap_{j\in\mathbb Z}\mathcal F_j$, and define the projection operators $\mathcal P_jY:=\E(Y\mid\mathcal F_j)-\E(Y\mid\mathcal F_{j-1})$ for integrable $Y$.

\begin{lemma}[Dependence transfer to lag-score vectors]
\label{lem:dependence-transfer}
Under Assumption~\ref{asm:data-score}, a constant \(C\) independent of \(n,t,j,k\) satisfies:
\begin{enumerate}
\item[(i)] For \(j\ge0\) and \(1\le k\le K_n\), \(\|\widetilde W_{t,n,k}-\widetilde W_{t,n,k}^{*(j)}\|_q\le C\{\delta_{j,q(1+\gamma)}^X+\1\{j\ge k\}\delta_{j-k,q(1+\gamma)}^X\}\), with \(\delta_{u,q(1+\gamma)}^X=0\) for \(u<0\).
\item[(ii)] The moments satisfy \(\sup_{n,t,k}\|\widetilde W_{t,n,k}\|_q\le C\).
\end{enumerate}
\end{lemma}

\begin{lemma}[Uniform partial-sum moment bound]
\label{lem:app-partial-sum-moment}
Under Assumption~\ref{asm:data-score}, there exists a constant $C$ independent of $n$, $k$, and $I$ such that
\[
\left\| \sum_{t\in I}\widetilde W_{t,n,k} \right\|_q \le C|I|^{1/2}
\]
for all $n$, all $1\le k\le K_n$, and every finite interval $I\subset\mathbb Z$. The corresponding maximal bound is \(\|\max_{1\le m\le M}|\sum_{t=a+1}^{a+m}\widetilde W_{t,n,k}|\|_q\le C\sqrt M\) for every integer \(a\) and positive integer \(M\). Both bounds hold for the forward scores \(h(X_t,X_{t+k})-\theta_k\) as well.
\end{lemma}

\begin{lemma}[Absolute and weighted summability of lag-score covariances]
\label{lem:app-covariance-summability}
Under Assumption~\ref{asm:data-score}, there exists a constant $C$ independent of $n$, $k$, and $\ell$ such that, for all $n$ and $1\le k,\ell\le K_n$,
\[
\sum_{r\in\mathbb Z} \left| \Cov(\widetilde W_{0,n,k},\widetilde W_{r,n,\ell}) \right| \le C
\]
and
\[
\sum_{r\in\mathbb Z}|r| \left| \Cov(\widetilde W_{0,n,k},\widetilde W_{r,n,\ell}) \right| \le C K_n.
\]
Moreover, for the centered forward scores $W_{t,n}:=H_{t,n}-\Theta_n$, the same bounds hold:
\[
\sum_{r\in\mathbb Z} \left|\Cov(W_{0,n,k},W_{r,n,\ell})\right| \le C, \qquad \sum_{r\in\mathbb Z}|r| \left|\Cov(W_{0,n,k},W_{r,n,\ell})\right| \le C K_n.
\]
Consequently, $\Sigma_n$ in \eqref{eq:common-long-run-covariance} is well defined, positive semidefinite, and satisfies $\sup_n\|\Sigma_n\|_{\max}<\infty$.

\end{lemma}

We use a finite-memory approximation of the centered scores to construct independent blocks.

\begin{lemma}[Finite-memory approximation of the scores]
\label{lem:app-blocking-mdep}
Suppose Assumption~\ref{asm:data-score} holds. Write $W_{t,n}=H_{t,n}-\Theta_n$ and, for an integer $m\ge1$, define $V_{t,n}^{(m)}:=\E(W_{t,n}\mid\varepsilon_{t-m+1},\ldots,\varepsilon_{t+K_n})$ and $M=K_n+m$. The following bounds hold with a constant $C$ uniform in $n$ and all indices and lengths specified below:
\begin{enumerate}
\item[(i)] The sequence $\{V_{t,n}^{(m)}\}$ is stationary and centered. Collections indexed by sets separated by at least $M$ are independent.
\item[(ii)] For every integer $a$, every positive integer $L$, and every lag coordinate $1\le k\le K_n$,
\[
\left\|\max_{1\le u\le L}\left|\sum_{t=a+1}^{a+u}(W_{t,n,k}-V_{t,n,k}^{(m)})\right|\right\|_q\le C\sqrt L\,m^{1-\beta}.
\]
\item[(iii)] For the same indices, $\|\max_{1\le u\le L}|\sum_{t=a+1}^{a+u}V_{t,n,k}^{(m)}|\|_q\le C\sqrt L$.
\item[(iv)] For every integer $r\ge1$, the matrix $\Sigma_n^{(r,m)}:=r^{-1}\Cov(\sum_{t=1}^rV_{t,n}^{(m)})$ satisfies $\|\Sigma_n^{(r,m)}-\Sigma_n\|_{\max}\le C(K_n/r+m^{1-\beta})$.
\end{enumerate}
\end{lemma}

\begin{lemma}[Partial-sum maxima under blocking]
\label{lem:app-blocking-fluctuation}
Under Assumptions~\ref{asm:data-score} and \ref{asm:block-complexity}(i), there is a constant $C$ such that, for all sufficiently large $n$,
\[
\E\max_{0\le u\le N_n}\left\|N_n^{-1/2}\sum_{t=1}^u(H_{t,n}-\Theta_n)\right\|_\infty\le C\ell_n^{1/2},
\]
where $\ell_n=\log(2N_nK_n+2d_n^{\max})$ is defined in Assumption~\ref{asm:block-complexity}.
\end{lemma}

\begin{lemma}[Gaussian anti-concentration and comparison]
\label{lem:app-anti-concentration}
Let \(Z_n=(Z_{n,1},\ldots,Z_{n,p_n})^\top\) be a centered Gaussian vector and \(M_n=\max_{j\le p_n}|Z_{n,j}|\).
\begin{enumerate}
\item[(i)] If \(\inf_n\min_{j\le p_n}\Var(Z_{n,j})\ge\underline\sigma^2>0\), then \(\sup_u\mathbb{P}(|M_n-u|\le t)\le Ct\{1\vee\sqrt{\log(2p_n)}\}\) for every \(t>0\), where \(C\) depends only on \(\underline\sigma\).
\item[(ii)] If, in addition, \(\sup_n\max_{j\le p_n}\Var(Z_{n,j})\le\overline\sigma^2<\infty\), and \(Z_n'\) is another centered Gaussian vector with \(\Delta_n=\|\Cov(Z_n')-\Cov(Z_n)\|_{\max}\), then \(\sup_u|\mathbb{P}(\|Z_n'\|_\infty\le u)-\mathbb{P}(M_n\le u)|\le C\Delta_n^{1/3}\{1\vee\log(2p_n/\Delta_n)\}^{2/3}\), where \(C\) depends only on \(\underline\sigma\) and \(\overline\sigma\). The expression is defined as zero when \(\Delta_n=0\).
\end{enumerate}
\end{lemma}

\begin{lemma}[Calibration by a common Gaussian target]
\label{lem:app-quantile-calibration}
Let $F_n$ be the distribution function of $\|Z_n\|_\infty$, where $Z_n$ is a centered Gaussian vector with at least one coordinate of positive variance. Suppose \(\sup_u|\mathbb{P}(T_n\le u)-F_n(u)|\to0\) and \(\sup_u|\mathbb{P}^*(T_n^*\le u)-F_n(u)|\xrightarrow{p}0\). For \(c_n^*:=\inf\{u:\mathbb{P}^*(T_n^*\le u)\ge1-\alpha\}\), with fixed \(\alpha\in(0,1)\), we have \(\mathbb{P}(T_n>c_n^*)\to\alpha\).
\end{lemma}

For a deterministic terminal transformation \(\mathsf A_n\) with \(d_{A,n}\) rows, define
\[
r_{n,A}^{\mathrm{boot}}:=\left\|\mathsf A_n\{\Cov^*(U_n^*)-\Sigma_n\}\mathsf A_n^\top\right\|_{\max}.
\]
For the temporal scan, with \(p_n=|\mathcal S_n(\epsilon)|K_n\) as in the main text, define
\[
r_{n,\mathrm{br}}^{\mathrm{boot}}:=\sup_{s,s'\in\mathcal S_n(\epsilon)}\left\|\Cov^*\{B_n^*(s),B_n^*(s')\}-\left\{\frac{\min(\lfloor N_ns\rfloor,\lfloor N_ns'\rfloor)}{N_n}-\frac{\lfloor N_ns\rfloor\lfloor N_ns'\rfloor}{N_n^2}\right\}\Sigma_n\right\|_{\max}.
\]
The target covariance is that of \(G_n^{\mathrm{br}}\). Define the empirical covariance kernel
\[
\widehat C_n(s,s'):=\frac1{N_n}\sum_{\ell=1}^{L_n}\left\{\sum_{\substack{t\in I_{\ell,n}\\
t\le\lfloor N_ns\rfloor}}(H_{t,n}-\widehat\Theta_n)\right\}\left\{\sum_{\substack{t\in I_{\ell,n}\\
t\le\lfloor N_ns'\rfloor}}(H_{t,n}-\widehat\Theta_n)\right\}^{\top},\qquad s,s'\in[0,1].
\]
Its Gaussian target is \(C_n^G(s,s'):=\lfloor N_n(s\wedge s')\rfloor\Sigma_n/N_n\). At \((1,1)\), the feasible kernel equals \(\Cov^*(U_n^*)\); applying the bridge operation to both arguments gives the covariance kernel of \(B_n^*\).

\begin{lemma}[Primitive block-covariance bound]
\label{lem:app-block-covariance}
Suppose Assumption~\ref{asm:data-score} holds and $b_n$ is a positive integer sequence satisfying $b_n=o(N_n)$. With $\mathfrak b_n$ denoting the population block-covariance bias defined in the main text, set $\rho_n:=\mathfrak b_n+K_n^{4/q}(b_n/N_n)^{1/2}$. Then:
\begin{enumerate}
\item[(i)] Uniformly over the endpoints, \(\sup_{s,s'\in[0,1]}\|\widehat C_n(s,s')-C_n^G(s,s')\|_{\max}=O_p(\rho_n)\).
\item[(ii)] For deterministic terminal transformations with uniformly bounded row \(\ell_1\)-norms, \(r_{n,A}^{\mathrm{boot}}=O_p(\rho_n)\). For the temporal bridge, \(r_{n,\mathrm{br}}^{\mathrm{boot}}=O_p(\rho_n)\).
\item[(iii)] If Assumption~\ref{asm:block-complexity}(ii) also holds, then \(\{r_{n,A}^{\mathrm{boot}}\}^{1/3}\{1\vee\log(2d_{A,n}/r_{n,A}^{\mathrm{boot}})\}^{2/3}\xrightarrow{p}0\) for every terminal transformation under consideration, and \(\{r_{n,\mathrm{br}}^{\mathrm{boot}}\}^{1/3}\{1\vee\log(2p_n/r_{n,\mathrm{br}}^{\mathrm{boot}})\}^{2/3}\xrightarrow{p}0\) for the temporal scan. Each expression is defined as zero when its covariance error is zero.
\end{enumerate}
\end{lemma}

\begin{lemma}[Sampling and bootstrap moments for a fixed lag horizon]
\label{lem:app-fixed-window-moments}
Suppose Assumption~\ref{asm:data-score} holds with fixed \(K_n\equiv K\). Then \(\E\|U_n\|_2^2\le C\) and \(\E\E^*\|U_n^*\|_2^2\le C(1+b_n/N_n)\). Consequently, when \(b_n=o(N_n)\), \(U_n=O_p(1)\), \(\widehat\Theta_n-\Theta_n=O_p(N_n^{-1/2})\), and \(U_n^*=O_p^*(1)\) in probability.
\end{lemma}

\begin{lemma}[Deterministic bridge drift and crossing-pair error]
\label{lem:app-cp-drift}
Under the single-change model of Section~\ref{subsec:theory-cp}, suppose the within-regime and crossing-pair score means are uniformly bounded. Then \(\sup_{s\in\mathcal S_n(\epsilon)}\|\E B_n(s)-\sqrt{N_n}\mathcal D_n(s)\|_\infty\le CK_n/\sqrt{N_n}\).
\end{lemma}

\begin{lemma}[Exact curvature expansion]
\label{lem:curvature-taylor}
For fixed \(K_n\equiv K\ge3\) and every \(\theta,e\in\mathbb R^K\), the following identities and bounds hold:
\begin{enumerate}
\item[(i)] \(\Phi_n(\theta+e)=\Phi_n(\theta)+D\Phi_n(\theta)e+\mathcal R_n(e)\), where the \(k\)-th remainder coordinate is \(\mathcal R_{n,k}(e)=e_ke_{k+2}-e_{k+1}^2\).
\item[(ii)] \(\|\mathcal R_n(e)\|_\infty\le2\|e\|_\infty^2\).
\item[(iii)] \(\|D\Phi_n(\theta+e)-D\Phi_n(\theta)\|_{\infty\to\infty}\le4\|e\|_\infty\), and \(\|D\Phi_n(\theta)\|_{\infty\to\infty}\le4\|\theta\|_\infty\).
\end{enumerate}
\end{lemma}

\subsection{Verification for simulation constructions}
\label{subsec:supp-model-verification}

For a finite-dimensional Gaussian linear process \(X_t=\sum_{j\ge0}\Psi_j\varepsilon_{t-j}\) with \(\|\Psi_j\|_{\mathrm{op}}\le C\rho^j\), \(0<\rho<1\), replacing \(\varepsilon_0\) gives \(\|X_j-X_j^*\|_u\le C\rho^j\|\varepsilon_0-\varepsilon_0'\|_u\) for every finite \(u\). The Gaussian MA and stable AR designs therefore satisfy Assumption~\ref{asm:data-score}(i),(iii) at every finite moment order and every fixed polynomial exponent \(\beta\). Distance-based pairwise functions use \(\gamma=0\), and the unbounded Euclidean inner product uses \(\gamma=1\); the latter calls for observation moments of order \(2q\), while all subsequent score bounds remain of order \(q\). The inner-product bound follows from \(|\langle x,y\rangle-\langle x',y'\rangle|\le\|x-x'\|\|y\|+\|x'\|\|y-y'\|\).

The mixture-density map in Section~\ref{subsec:simulation-range} also preserves geometric physical dependence in the Hellinger metric. For fixed component densities, its Hellinger distance is bounded by the Hellinger distance between the Bernoulli mixing weights. The map \(z\mapsto(\sqrt{\pi_c(z)},\sqrt{1-\pi_c(z)})\) has bounded derivative, so \(d_H(\mathcal G_c(z),\mathcal G_c(z'))\le C|z-z'|\). The square-root-density map in the first temporal design is Lipschitz in the latent Gaussian vector because \(a_t\) stays below \(1/\sqrt3\). Both constructions are bounded in Hellinger distance and satisfy the required metric moment condition. For the bounded density design in Section~\ref{subsec:simulation-autoregression}, each symmetric two-state chain with correlation parameter \(0<\rho<1\) admits a refresh representation: at each step it retains its state with probability \(\rho\) and otherwise draws an independent symmetric sign. A perturbation survives \(j\) steps with probability at most \(\rho^j\), giving geometric physical dependence in every finite moment order. The finite collection of chains and bounded linear density map preserve this property. These constructions satisfy Assumption~\ref{asm:data-score} within each stationary regime.

For the Euclidean inner-product design, Assumption~\ref{asm:covariance-nondegeneracy} follows from the spectral representation. Let \(f_X\) denote the spectral density of one scalar coordinate, normalized so that its autocovariance is \((2\pi)^{-1}\int_{-\pi}^{\pi}e^{ik\lambda}f_X(\lambda)\,d\lambda\). Gaussian fourth-moment expansion and independence of the \(p\) coordinate series give, for \(a\in\mathbb R^K\),
\[
a^\top\Sigma a=\frac{2}{p}\frac1{2\pi}\int_{-\pi}^{\pi}f_X(\lambda)^2\left\{\sum_{k=1}^K a_k\cos(k\lambda)\right\}^2d\lambda\ge\frac{\inf_\lambda f_X(\lambda)^2}{p}\|a\|_2^2.
\]
The equality follows by summing \(\Cov(X_0X_k,X_rX_{r+\ell})=\gamma_X(r)\gamma_X(r+\ell-k)+\gamma_X(r+\ell)\gamma_X(r-k)\) over \(r\); the inequality uses orthogonality of the cosine functions. The stable AR(2) parameter range in the simulation has a spectral density bounded away from zero, and its curvature Jacobian rows are nonzero for the fixed \(K=8\). Since the simulated parameter set is finite, the lower bound is uniform over that set. For the remaining designs, Assumption~\ref{asm:covariance-nondegeneracy} is imposed separately.

For the fixed lag horizons used in the simulations, the block-length rule gives $b_n\asymp N_n^{1/3}$. Lemma~\ref{lem:app-covariance-summability} yields $\mathfrak b_n=O(b_n^{-1})$, so Assumption~\ref{asm:block-complexity}(ii) holds, including the logarithmic factor for the temporal scan. To verify part~(i), take $m_n=\lceil N_n^{1/4}\rceil$ and $r_n=\lceil N_n^{1/2}\rceil$. Since $\ell_n=O(\log N_n)$, the three expressions in part~(i) are respectively
\[
O\!\left(N_n^{-(\beta-1)/4}\log^2N_n\right),\qquad
O\!\left(N_n^{-1/4}\log^2N_n\right),\qquad
O\!\left(N_n^{-(1-2/q)/2}\log^7N_n\right),
\]
all of which tend to zero.
\section{Proofs of Theorems and Propositions}
\label{sec:supp-theorems}

\subsection{Proofs of the common approximation results}

\subsubsection{Proof of Theorem~\ref{thm:master-gaussian}}

\begin{proof}
Write $N=N_n$, $K=K_n$, $m=m_n$, $r=r_n$, $M=K+m$, $P=r+M$, and $\ell=\ell_n$. Let $W_t=H_{t,n}-\Theta_n$ and $V_t=V_{t,n}^{(m)}$ from Lemma~\ref{lem:app-blocking-mdep}. Assumption~\ref{asm:block-complexity}(i) implies $M/r\to0$ and $r/N\to0$.

\emph{Step 1: Block decomposition.}
Put $J=\lfloor N/P\rfloor$ and define
\[
I_j=\{(j-1)P+1,\ldots,(j-1)P+r\},\qquad I'_j=\{(j-1)P+r+1,\ldots,jP\},
\]
for $1\le j\le J$. Let $Y_j=\sum_{t\in I_j}V_t$, $Y'_j=\sum_{t\in I'_j}V_t$, and $R=\sum_{t=JP+1}^NV_t$. For all sufficiently large $n$, $r\ge M$ and $J\ge4$. Lemma~\ref{lem:app-blocking-mdep} makes the $Y_j$ independent and identically distributed, as are the $Y'_j$, and gives
\[
\|Y_{j,k}\|_q\le C\sqrt r,\qquad \|Y'_{j,k}\|_q\le C\sqrt M,\qquad \|R_k\|_q\le C\sqrt P.
\]
Write $\Sigma^{(r)}=\Sigma_n^{(r,m)}$. The same lemma gives $\|\Sigma^{(r)}-\Sigma_n\|_{\max}\le C(K/r+m^{1-\beta})$.

\emph{Step 2: Bridge approximation.}
For $s\in\mathcal S_n(\epsilon)$, set $\lambda_s=\lfloor Ns\rfloor/N$, $j_s=\lfloor N\lambda_s/P\rfloor$, and $c_{sj}=\1\{j\le j_s\}-\lambda_s$. Define the large-block bridge by
\[
\Xi_n(s)=N^{-1/2}\sum_{j=1}^Jc_{sj}Y_j.
\]
Replacing $W_t$ by $V_t$ in $B_n(s)$ incurs expected uniform error at most $CK^{1/q}m^{1-\beta}$ by Lemma~\ref{lem:app-blocking-mdep}(ii). The resulting bridge differs from $\Xi_n(s)$ by
\[
N^{-1/2}\left\{\sum_{j\le j_s}Y'_j-\lambda_s\sum_{j=1}^JY'_j+\sum_{t=j_sP+1}^{\lfloor Ns\rfloor}V_t-\lambda_sR\right\}.
\]
Doob's and Burkholder's inequalities bound the expected uniform gap contribution by $CK^{1/q}(M/r)^{1/2}$. Lemma~\ref{lem:app-blocking-mdep}(iii), applied to the at most $J+1$ periods of length $P$, bounds the unfinished-period contribution by $CK^{1/q}(P/N)^{1/2-1/q}$. The remainder contributes at most $CK^{1/q}(P/N)^{1/2}$. Consequently,
\[
\E\sup_{s\in\mathcal S_n(\epsilon)}\|B_n(s)-\Xi_n(s)\|_\infty\le CK^{1/q}\left\{m^{1-\beta}+\left(\frac Mr\right)^{1/2}+\left(\frac PN\right)^{1/2-1/q}\right\}=o(\ell^{-1/2}).
\]

\emph{Step 3: Gaussian approximation and comparison.}
Independence of the large blocks gives
\[
\Cov\{\Xi_n(s),\Xi_n(s')\}=\frac rN\{j_s\wedge j_{s'}-\lambda_{s'}j_s-\lambda_sj_{s'}+J\lambda_s\lambda_{s'}\}\Sigma^{(r)}.
\]
Since $|j_sP/N-\lambda_s|\le P/N$ and $|JP/N-1|\le P/N$, Step~1 and Lemma~\ref{lem:app-covariance-summability} imply
\[
\sup_{s,s'\in\mathcal S_n(\epsilon)}\left\|\Cov\{\Xi_n(s),\Xi_n(s')\}-(\lambda_s\wedge\lambda_{s'}-\lambda_s\lambda_{s'})\Sigma_n\right\|_{\max}\le C\Delta,
\]
where $\Delta=(K+M)/r+m^{1-\beta}+P/N$ satisfies $\Delta\ell^2\to0$. Trimming and Assumption~\ref{asm:covariance-nondegeneracy} therefore bound the coordinate variances of $\Xi_n(s)$ away from zero for all sufficiently large $n$.

Consider the independent centered vectors with coordinates $X_{j,(s,k)}=(J/N)^{1/2}c_{sj}Y_{j,k}$. Their normalized sum has coordinates $\Xi_{n,k}(s)$, and their averaged coordinate variances are bounded away from zero. Since $q\ge4$, $|c_{sj}|\le1$, and $Jr/N\le1$,
\[
\frac1J\sum_{j=1}^J\E|X_{j,(s,k)}|^{2+\kappa}\le C,\quad \kappa=1,2,\qquad \E\max_{s,k}|X_{j,(s,k)}|^q\le CK.
\]
Let $\Xi_n^G$ be a centered Gaussian array with the covariance of $\Xi_n$, and put $T^\Xi=\sup_s\|\Xi_n(s)\|_\infty$ and $T^{\Xi,G}=\sup_s\|\Xi_n^G(s)\|_\infty$, with suprema over $\mathcal S_n(\epsilon)$. Applying \citet[Proposition~2.1]{chernozhukov2017central} with envelope $CK^{1/q}$ gives
\[
\sup_u|\mathbb P(T^\Xi\le u)-\mathbb P(T^{\Xi,G}\le u)|\le C\left\{\left(\frac{K^{2/q}\ell^7r}{N}\right)^{1/6}+\left[K^{2/q}\ell^3\left(\frac rN\right)^{1-2/q}\right]^{1/3}\right\}=o(1).
\]
Here $J^{-1}\le4r/N$ and $\log\{(p_n\vee3)J\}\le C\ell$; coordinates may be duplicated when the dimension is below three.

Let $T=\sup_{s\in\mathcal S_n(\epsilon)}\|B_n(s)\|_\infty$ and $T^G=\sup_{s\in\mathcal S_n(\epsilon)}\|G_n^{\mathrm{br}}(s)\|_\infty$. The covariance bound above and Lemma~\ref{lem:app-anti-concentration}(ii) yield
\[
\sup_u|\mathbb P(T^{\Xi,G}\le u)-\mathbb P(T^G\le u)|\le C\Delta^{1/3}\left(\ell+\log\frac1\Delta\right)^{2/3}=o(1).
\]
By Step~2, choose a deterministic $t_n>0$ such that $\E|T-T^\Xi|=o(t_n)$ and $t_n\sqrt\ell\to0$. Markov's inequality and Lemma~\ref{lem:app-anti-concentration}(i) then give
\[
\sup_u|\mathbb P(T\le u)-\mathbb P(T^G\le u)|\le o(1)+\frac{\E|T-T^\Xi|}{t_n}+Ct_n\sqrt\ell=o(1).
\]

\emph{Step 4: Terminal transformations.}
Write the rows of $\mathsf A_n$ as $a_v^\top$. Their uniformly bounded $\ell_1$-norms and the block bounds give
\[
\E\left|\|\mathsf A_nU_n\|_\infty-\left\|N^{-1/2}\sum_{j=1}^J\mathsf A_nY_j\right\|_\infty\right|\le CK^{1/q}\left\{m^{1-\beta}+\left(\frac Mr\right)^{1/2}+\left(\frac PN\right)^{1/2}\right\}=o(\ell^{-1/2}).
\]
For $X_{j,v}=(J/N)^{1/2}a_v^\top Y_j$, the averaged variances are $(Jr/N)a_v^\top\Sigma^{(r)}a_v$ and are bounded away from zero by Assumption~\ref{asm:covariance-nondegeneracy} and Step~1. The row-norm bound gives uniformly bounded third and fourth moments and $\E\max_v|X_{j,v}|^q\le CK$. Thus the same independent-vector approximation applies. The resulting Gaussian covariance is $(Jr/N)\mathsf A_n\Sigma^{(r)}\mathsf A_n^\top$, whose entrywise distance from $\mathsf A_n\Sigma_n\mathsf A_n^\top$ is at most $C\Delta$. Gaussian comparison and smoothing as in Step~3 prove the terminal conclusion.
\end{proof}
\subsubsection{Proof of Theorem~\ref{thm:master-bootstrap}}

\begin{proof}
Let \(\rho_n\) be as in Lemma~\ref{lem:app-block-covariance}. Under Assumption~\ref{asm:block-complexity}(ii), that lemma gives \(r_{n,A}^{\mathrm{boot}}=O_p(\rho_n)\) and \(r_{n,\mathrm{br}}^{\mathrm{boot}}=O_p(\rho_n)\), with both Gaussian comparison bounds tending to zero in probability. Conditional on the observations, the bootstrap vectors are centered Gaussian.
\begin{enumerate}
\item[(i)] Let \(F_{n,A}^G(u)=\mathbb{P}\{\|\mathsf A_nG_n(1)\|_\infty\le u\}\). Lemma~\ref{lem:app-covariance-summability} gives \(\|\Sigma_n\|_{\max}\le C\). For every row \(a^\top\) of \(\mathsf A_n\), the row-norm bound and Assumption~\ref{asm:covariance-nondegeneracy} therefore give \(\underline\sigma^2\le a^\top\Sigma_na\le\|a\|_1^2\|\Sigma_n\|_{\max}\le C\). The conditional covariance discrepancy is exactly \(r_{n,A}^{\mathrm{boot}}\). Lemma~\ref{lem:app-anti-concentration}(ii) yields
\[
\sup_u\left|\mathbb{P}^*\{\|\mathsf A_nU_n^*\|_\infty\le u\}-F_{n,A}^G(u)\right|\le C\{r_{n,A}^{\mathrm{boot}}\}^{1/3}\{1\vee\log(2d_{A,n}/r_{n,A}^{\mathrm{boot}})\}^{2/3}=o_p(1).
\]
\item[(ii)] Write \(\lambda_s=\lfloor N_ns\rfloor/N_n\), and stack the \(p_n\) lag--time coordinates of the bootstrap bridge and of its Gaussian target. The covariance discrepancy is \(r_{n,\mathrm{br}}^{\mathrm{boot}}\), and the target variances satisfy \(c_\epsilon\underline\sigma^2\le\lambda_s(1-\lambda_s)(\Sigma_n)_{kk}\le C\). With \(F_{n,\mathrm{br}}^G(u)=\mathbb{P}\{\sup_{s\in\mathcal S_n(\epsilon)}\|G_n^{\mathrm{br}}(s)\|_\infty\le u\}\), the same comparison gives
\[
\sup_u\left|\mathbb{P}^*\{\sup_{s\in\mathcal S_n(\epsilon)}\|B_n^*(s)\|_\infty\le u\}-F_{n,\mathrm{br}}^G(u)\right|\le C\{r_{n,\mathrm{br}}^{\mathrm{boot}}\}^{1/3}\{1\vee\log(2p_n/r_{n,\mathrm{br}}^{\mathrm{boot}})\}^{2/3}=o_p(1).
\]
\end{enumerate}
\end{proof}

\subsection{Profile stabilization}
\label{subsec:app-stabilization}

\subsubsection{Proof of Proposition~\ref{prop:stab-pointwise}}

\begin{proof}
Set \(\mathsf A_n=C_{m_{q_0}}H_{m_{q_0}}P_{q_0}\). Under \(H_0^{\mathrm{stab}}(q_0)\), \(\mathsf A_n\Theta_n=0\), so \(\sqrt{N_n}\mathsf A_n\widehat\Theta_n=\mathsf A_nU_n\), \(R_n(q_0)=\|\mathsf A_nU_n\|_\infty\), and \(R_n^*(q_0)=\|\mathsf A_nU_n^*\|_\infty\). The rows of \(\mathsf A_n\) belong to \(\mathcal T_n\) and have bounded \(\ell_1\)-norms because \(K\) and \(q_0\) are fixed. Let \(F_n(u)=\mathbb{P}\{\|\mathsf A_nG_n(1)\|_\infty\le u\}\). The terminal conclusions of Theorems~\ref{thm:master-gaussian} and \ref{thm:master-bootstrap} imply
\[
\begin{aligned}
&\sup_u|\mathbb{P}^*\{R_n^*(q_0)\le u\}-\mathbb{P}\{R_n(q_0)\le u\}|\\
&\le\sup_u|\mathbb{P}^*\{R_n^*(q_0)\le u\}-F_n(u)|\\
&\quad+\sup_u|\mathbb{P}\{R_n(q_0)\le u\}-F_n(u)|=o_p(1).
\end{aligned}
\]
Assumption~\ref{asm:covariance-nondegeneracy} and Lemma~\ref{lem:app-quantile-calibration} yield \(\mathbb{P}\{R_n(q_0)>\widehat c_{n,1-\alpha}^{\mathrm{stab}}(q_0)\}\to\alpha\).
\end{proof}

\subsubsection{Proof of Proposition~\ref{prop:stab-power}}

\begin{proof}
Set \(\mathsf A_n=C_{m_{q_0}}H_{m_{q_0}}P_{q_0}\) and \(D_n=\sqrt{N_n}\|\mathsf A_n\Theta_n\|_\infty\to\infty\). The exact identity \(\sqrt{N_n}\mathsf A_n\widehat\Theta_n=\sqrt{N_n}\mathsf A_n\Theta_n+\mathsf A_nU_n\) gives \(R_n(q_0)\ge D_n-\|\mathsf A_nU_n\|_\infty\). Lemma~\ref{lem:app-fixed-window-moments} and the bounded row norms yield \(\E\|\mathsf A_nU_n\|_\infty^2\le C\) and \(\E^*\{R_n^*(q_0)\}^2\le\|\mathsf A_n\|_{\infty\to\infty}^2\E^*\|U_n^*\|_2^2=O_p(1)\). Conditional Markov's inequality thus gives \(\widehat c_{n,1-\alpha}^{\mathrm{stab}}(q_0)\le[\E^*\{R_n^*(q_0)\}^2/\alpha]^{1/2}=O_p(1)\). Consequently,
\[
\mathbb{P}\{R_n(q_0)\le\widehat c_{n,1-\alpha}^{\mathrm{stab}}(q_0)\}\le\mathbb{P}\{\|\mathsf A_nU_n\|_\infty>D_n/2\}+\mathbb{P}\{\widehat c_{n,1-\alpha}^{\mathrm{stab}}(q_0)\ge D_n/2\}\le\frac{4C}{D_n^2}+o(1)\to0.
\]
\(C_mH_mv=0\) implies that all first \(m-1\) cumulative sums of \(H_mv\) vanish. Successive differences and \(\mathbf 1_m^\top H_mv=0\) then give \(H_mv=0\), which is equivalent to \(v\) being constant. Thus every fixed nonconstant tail has \(\|\mathsf A_n\Theta_n\|_\infty>0\).
\end{proof}

\subsubsection{Proof of Proposition~\ref{prop:stab-cutoff}}

\begin{proof}

Suppose $q_\star<\infty$. For each candidate $q_0<q_\star$, Proposition~\ref{prop:stab-power} and the assumed signal condition give
\[
  \mathbb P\left\{R_n(q_0)\le\widehat c_{n,1-\alpha}^{\mathrm{stab}}(q_0)\right\}\to0.
\]
Let $E_n$ be the event that every candidate $q_0<q_\star$ is rejected, with $E_n$ equal to the whole sample space when $q_\star=0$. Since $K$ is fixed, a finite union bound gives $\mathbb P(E_n^c)\to0$. Consequently, $\mathbb P(\widehat q_{n,\alpha}<q_\star)\le\mathbb P(E_n^c)\to0$.

On $E_n$, the estimator equals $q_\star$ if and only if the test at $q_\star$ does not reject. Therefore
\[
  \left|\mathbb P(\widehat q_{n,\alpha}=q_\star)-\mathbb P\left\{R_n(q_\star)\le\widehat c_{n,1-\alpha}^{\mathrm{stab}}(q_\star)\right\}\right|
  \le\mathbb P(E_n^c)\to0.
\]
Since $H_0^{\mathrm{stab}}(q_\star)$ holds, Proposition~\ref{prop:stab-pointwise} makes the second probability tend to $1-\alpha$, proving the finite-cutoff assertions.

If $q_\star=\infty$, every candidate satisfies the alternative and the assumed signal condition. Proposition~\ref{prop:stab-power} and a finite union bound yield
\[
  \mathbb P(\widehat q_{n,\alpha}<\infty)
  \le\sum_{q_0=0}^{K-2}\mathbb P\left\{R_n(q_0)\le\widehat c_{n,1-\alpha}^{\mathrm{stab}}(q_0)\right\}\to0.
\]
Thus $\mathbb P(\widehat q_{n,\alpha}=\infty)\to1$.
\end{proof}
\subsection{Temporal stability}
\label{subsec:app-temporal-change}

\subsubsection{Proof of Proposition~\ref{prop:cp-bootstrap}}

\begin{proof}
Under Assumption~\ref{asm:data-score} and \(H_0^{\mathrm{temp}}\), the lag-score sequence is stationary with common mean \(\Theta_n\). Let \(F_n(u)=\mathbb{P}\{\sup_{s\in\mathcal S_n(\epsilon)}\|G_n^{\mathrm{br}}(s)\|_\infty\le u\}\). Since the observed and bootstrap statistics are exactly the bridge maxima, the bridge conclusions of Theorems~\ref{thm:master-gaussian} and \ref{thm:master-bootstrap} give
\[
\begin{aligned}
&\sup_u|\mathbb{P}^*\{T_n^*(K_n,\epsilon)\le u\}-\mathbb{P}\{T_n(K_n,\epsilon)\le u\}|\\
&\le\sup_u|\mathbb{P}^*\{T_n^*(K_n,\epsilon)\le u\}-F_n(u)|\\
&\quad+\sup_u|\mathbb{P}\{T_n(K_n,\epsilon)\le u\}-F_n(u)|=o_p(1).
\end{aligned}
\]
Trimming and Assumption~\ref{asm:covariance-nondegeneracy} bound the scanned Gaussian variances away from zero. Lemma~\ref{lem:app-quantile-calibration} then gives \(\mathbb{P}\{T_n(K_n,\epsilon)>\widehat c_{n,1-\alpha}^{\mathrm{CP}}\}\to\alpha\).
\end{proof}

\subsubsection{Proof of Theorem~\ref{thm:cp-detection}}

\begin{proof}
Write $N=N_n$, $K=K_n$, $b=b_n$, $\Delta=\Delta_{n,h}$, and $s_{0,n}=m_0/N$. Assumption~\ref{asm:block-complexity}(ii), imposed within each regime, gives $b\to\infty$ and $b/N\to0$.

\emph{Step 1: Sampling fluctuation.}
For $r=1,2$, define $H_{t,n}^{(r)}=\bigl(h(X_t^{(r)},X_{t+1}^{(r)}),\ldots,h(X_t^{(r)},X_{t+K}^{(r)})\bigr)^\top$ and $W_{t,n}^{(r)}=H_{t,n}^{(r)}-\Theta_n^{(r)}$. Lemma~\ref{lem:app-blocking-fluctuation}, applied within each regime, and $\ell_n\asymp\log p_n$ give
\[
\max_{r=1,2}\max_{0\le m\le N}\left\|N^{-1/2}\sum_{t=1}^mW_{t,n}^{(r)}\right\|_\infty=O_p(\sqrt{\log p_n}).
\]
Let $\check H_{t,n}$ equal $H_{t,n}^{(1)}$ for $t\le m_0$ and $H_{t,n}^{(2)}$ otherwise, and define $\check W_{t,n}$ by the same splice. Each partial sum of $\check W_{t,n}$ is a combination of at most three regime partial sums, so the same bound holds for the spliced sequence.

Put $R_{t,n}=H_{t,n}-\check H_{t,n}$. Its $k$th coordinate vanishes unless $t\le m_0<t+k$. The growth-Lipschitz condition and the uniform metric moments imply
\[
M_{R,n}:=\max_{k\le K}\sum_{t=1}^N|R_{t,n,k}|\le C\sum_{v=m_0-K+1}^{m_0+K}\left\{1+d(X_v^{(1)},x_0)^{1+\gamma}+d(X_v^{(2)},x_0)^{1+\gamma}\right\},
\]
with $\E M_{R,n}\le CK$. The bridge of $R_{t,n}-\E R_{t,n}$ is bounded uniformly by $2N^{-1/2}(M_{R,n}+\E M_{R,n})=O_p(K/\sqrt N)$. Combining the two contributions gives
\begin{equation}
\sup_{s\in\mathcal S_n(\epsilon)}\|B_n(s)-\E B_n(s)\|_\infty=O_p\left(\sqrt{\log p_n}+\frac{K_n}{\sqrt{N_n}}\right).
\label{eq:app-cp-fluctuation}
\end{equation}

\emph{Step 2: Bootstrap critical value.}
For a scalar or vector sequence $v=(v_t)_{t=1}^N$, define
\[
V_\xi(v;m)=N^{-1/2}\sum_{\ell=1}^{L_n}\xi_\ell\sum_{\substack{t\in I_{\ell,n}\\t\le m}}v_t,\qquad \mathcal B_\xi(v;m)=V_\xi(v;m)-\frac mN V_\xi(v;N).
\]
Let $\E_\xi$ denote expectation over the multipliers conditional on both regime processes. For the observed statistic, $\E_\xi T_n^*=\E^*T_n^*$.

For $v_t\equiv1$ and $v_t=a_t:=\1\{t\le m_0\}-s_{0,n}$,
\[
\E_\xi\max_m|V_\xi(v;m)|+\E_\xi\max_m|\mathcal B_\xi(v;m)|\le C\sqrt b.
\]
Indeed, at block endpoints $V_\xi(v;m)$ is a Gaussian martingale with terminal variance at most $b$. Doob's inequality controls its maximum there. Within each block the path is linear, except possibly at $m_0$, whose additional contribution is at most $b\E|\xi_\ell|/\sqrt N\le C\sqrt b$.

For $r=1,2$, let $\bar W_n^{(r)}=N^{-1}\sum_{t=1}^NW_{t,n}^{(r)}$ and $Q_{t,n}^{(r)}=W_{t,n}^{(r)}-\bar W_n^{(r)}$. Lemmas~\ref{lem:app-block-covariance}(i) and \ref{lem:app-covariance-summability}, together with the within-regime covariance conditions, give
\[
\max_{r,m,k}\Var_\xi\{V_\xi(Q^{(r)};m)_k\}=O_p(1).
\]
The conditional Gaussian maximal inequality consequently yields $\E_\xi\max_{r,m}\|V_\xi(Q^{(r)};m)\|_\infty=O_p(\sqrt{\log p_n})$. Since $\max_r\|\bar W_n^{(r)}\|_\infty=O_p(\sqrt{\log p_n/N})$ by Step~1, the identity
\[
V_\xi(W^{(r)};m)=V_\xi(Q^{(r)};m)+\bar W_n^{(r)}V_\xi(1;m)
\]
gives the same bound for $W^{(r)}$. The splice identity also holds for $V_\xi$, because each block uses a common multiplier. Hence $\E_\xi\max_m\|\mathcal B_\xi(\check W;m)\|_\infty=O_p(\sqrt{\log p_n})$.

Write $\bar{\check W}_n=N^{-1}\sum_t\check W_{t,n}$ and $\bar R_n=N^{-1}\sum_tR_{t,n}$. Full-sample centering gives
\[
\begin{aligned}
B_n^*(m/N)={}&\mathcal B_\xi(\check W;m)-\bar{\check W}_n\mathcal B_\xi(1;m)+(\Theta_n^{(1)}-\Theta_n^{(2)})\mathcal B_\xi(a;m)\\
&+\mathcal B_\xi(R;m)-\bar R_n\mathcal B_\xi(1;m).
\end{aligned}
\]
Here $\|\bar{\check W}_n\|_\infty=O_p(\sqrt{\log p_n/N})$ and $\|\bar R_n\|_\infty\le M_{R,n}/N=O_p(K/N)$. Also,
\[
\E_\xi\max_m\|\mathcal B_\xi(R;m)\|_\infty\le\frac{2M_{R,n}}{\sqrt N}\E_\xi\max_\ell|\xi_\ell|=O_p\left\{\frac K{\sqrt N}\sqrt{\log(2L_n)}\right\}.
\]
Using $b/N\to0$ and $\log(2L_n)=O(\log p_n)$, these bounds give
\[
\E^*T_n^*(K_n,\epsilon)=O_p\left\{\left(1+\frac K{\sqrt N}\right)\sqrt{\log p_n}+\sqrt b\,\Delta\right\}.
\]
Conditional Markov's inequality therefore implies
\begin{equation}
\widehat c_{n,1-\alpha}^{\mathrm{CP}}\le\alpha^{-1}\E^*T_n^*(K_n,\epsilon)=O_p\left\{\left(1+\frac{K_n}{\sqrt{N_n}}\right)\sqrt{\log p_n}+\sqrt{b_n}\Delta_{n,h}\right\}.
\label{eq:app-cp-critical-bound}
\end{equation}

\emph{Step 3: Signal comparison.}
The metric moment and growth-Lipschitz conditions bound the within-regime and crossing-pair score means, so Lemma~\ref{lem:app-cp-drift} applies. At $s_{0,n}\in\mathcal S_n(\epsilon)$,
\[
\|\E B_n(s_{0,n})\|_\infty\ge c_\epsilon\sqrt{N_n}\Delta_{n,h}-\frac{CK_n}{\sqrt{N_n}}.
\]
By \eqref{eq:app-cp-fluctuation} and the signal condition,
\[
\frac{T_n(K_n,\epsilon)}{\sqrt{N_n}\Delta_{n,h}}\ge c_\epsilon-O_p\left(\frac{\sqrt{\log p_n}+K_n/\sqrt{N_n}}{\sqrt{N_n}\Delta_{n,h}}\right)=c_\epsilon-o_p(1).
\]
Meanwhile, \eqref{eq:app-cp-critical-bound} gives
\[
\frac{\widehat c_{n,1-\alpha}^{\mathrm{CP}}}{\sqrt{N_n}\Delta_{n,h}}=O_p\left\{\frac{(1+K_n/\sqrt{N_n})\sqrt{\log p_n}}{\sqrt{N_n}\Delta_{n,h}}+\sqrt{\frac{b_n}{N_n}}\right\}=o_p(1).
\]
The asserted rejection probability follows.
\end{proof}
\subsubsection{Proof of Theorem~\ref{thm:cp-localization}}

\begin{proof}
Let \(v_n=\sqrt{\log p_n}+K_n/\sqrt{N_n}\) and \(E_n=\sup_{s\in\mathcal S_n(\epsilon)}|\|B_n(s)\|_\infty-\sqrt{N_n}\|\mathcal D_n(s)\|_\infty|\). The norm inequality, Lemma~\ref{lem:app-cp-drift}, and \eqref{eq:app-cp-fluctuation} give \(E_n=O_p(v_n)\). The drift norm is Lipschitz with constant at most \(\Delta_{n,h}\). Thus, at \(s_{0,n}=m_0/N_n\), \(\|\mathcal D_n(s_{0,n})\|_\infty\ge\|\mathcal D_n(\tau_0)\|_\infty-\Delta_{n,h}/N_n\). Since \(\tau_0\) is interior and \(\eta_n\downarrow0\), the comparison set in \(\kappa_n(\eta_n)\) is nonempty for all sufficiently large \(n\); also \(|s_{0,n}-\tau_0|<\eta_n\) eventually by \(N_n\eta_n\to\infty\).

On the event \(|\widehat\tau_n-\tau_0|\ge\eta_n\), the definition of \(\kappa_n\) and maximality of \(\widehat\tau_n\) give
\[
0\le\|B_n(\widehat\tau_n)\|_\infty-\|B_n(s_{0,n})\|_\infty\le-\sqrt{N_n}\kappa_n(\eta_n)+\frac{\Delta_{n,h}}{\sqrt{N_n}}+2E_n.
\]
Uniform score moments imply \(\Delta_{n,h}\le C\), hence \(\Delta_{n,h}/\sqrt{N_n}=O(v_n)\). Condition~\eqref{eq:cp-localization-separation} therefore yields
\[
\mathbb{P}(|\widehat\tau_n-\tau_0|\ge\eta_n)\le\mathbb{P}\left\{\frac{2E_n+\Delta_{n,h}/\sqrt{N_n}}{v_n}\ge\frac{\sqrt{N_n}\kappa_n(\eta_n)}{v_n}\right\}\to0.
\]
In particular, \(|\widehat\tau_n-\tau_0|=O_p(\eta_n)\).
\end{proof}

\subsection{Geometric regularity}
\label{subsec:app-curvature}

\subsubsection{Proof of Proposition~\ref{prop:geometric-bootstrap}}

\begin{proof}
Write \(e_n=\widehat\Theta_n-\Theta_n\), \(J_n=D\Phi_n(\Theta_n)\), and \(\widehat J_n=D\Phi_n(\widehat\Theta_n)\). Assumption~\ref{asm:data-score} gives \(|\theta_k|\le |h(x_0,x_0)|+C\E\{1+d(X_0,x_0)\}^{1+\gamma}\le C\), uniformly in \(k\). Hence \(\|J_n\|_{\infty\to\infty}\le4\|\Theta_n\|_\infty\le C\). Lemma~\ref{lem:app-fixed-window-moments} gives \(\E\|e_n\|_\infty^2\le C/N_n\) and \(\E^*\|U_n^*\|_2^2=O_p(1)\). Let \(F_n(u)=\mathbb{P}\{\|J_nG_n(1)\|_\infty\le u\}\), and let \(\delta_n\) and \(\delta_n^*\) be the uniform distributional errors for \(\|J_nU_n\|_\infty\) and \(\|J_nU_n^*\|_\infty\), respectively, relative to \(F_n\). The terminal conclusions of Theorems~\ref{thm:master-gaussian} and \ref{thm:master-bootstrap} give \(\delta_n\to0\) and \(\delta_n^*\xrightarrow{p}0\). Since \(K\) is fixed and the Jacobian directions satisfy Assumption~\ref{asm:covariance-nondegeneracy}, Lemma~\ref{lem:app-anti-concentration}(i) gives \(\sup_u\mathbb{P}(|\|J_nG_n(1)\|_\infty-u|\le t)\le Ct\).
\begin{enumerate}
\item[(i)] Under \(H_0^{\mathrm{curv}}\), Lemma~\ref{lem:curvature-taylor} gives \(\sqrt{N_n}\Phi_n(\widehat\Theta_n)=J_nU_n+\sqrt{N_n}\mathcal R_n(e_n)\). Put \(D_n=\sqrt{N_n}\|\mathcal R_n(e_n)\|_\infty\le2\sqrt{N_n}\|e_n\|_\infty^2\). Then \(\E D_n\le C N_n^{-1/2}\). Smoothing with \(t_n=N_n^{-1/4}\) yields
\[
\sup_u|\mathbb{P}\{A_n(K_n)\le u\}-F_n(u)|\le\delta_n+\mathbb{P}(D_n>t_n)+Ct_n\le\delta_n+C N_n^{-1/4}\to0.
\]
\item[(ii)] The bootstrap plug-in discrepancy satisfies \(D_n^*:=\|\{\widehat J_n-J_n\}U_n^*\|_\infty\le4\|e_n\|_\infty\|U_n^*\|_2\). Conditional Markov's inequality gives \(\mathbb{P}^*(D_n^*>t_n)\le16t_n^{-2}\|e_n\|_\infty^2\E^*\|U_n^*\|_2^2=O_p(N_n^{-1/2})\). Therefore
\[
\sup_u|\mathbb{P}^*\{A_n^*(K_n)\le u\}-F_n(u)|\le\delta_n^*+\mathbb{P}^*(D_n^*>t_n)+Ct_n=o_p(1).
\]
Combining this with part~(i) by the triangle inequality proves the bootstrap approximation to the observed statistic.
\end{enumerate}
Lemma~\ref{lem:app-quantile-calibration} gives \(\mathbb{P}\{A_n(K_n)>\widehat c_{n,1-\alpha}^{\mathrm{curv}}(K_n)\}\to\alpha\).
\end{proof}

\subsubsection{Proof of Proposition~\ref{prop:geometric-power}}

\begin{proof}
Put \(e_n=\widehat\Theta_n-\Theta_n\) and \(D_n=\sqrt{N_n}\|\Phi_n(\Theta_n)\|_\infty\to\infty\). Lemma~\ref{lem:curvature-taylor} gives the exact decomposition \(\sqrt{N_n}\Phi_n(\widehat\Theta_n)=\sqrt{N_n}\Phi_n(\Theta_n)+D\Phi_n(\Theta_n)U_n+\sqrt{N_n}\mathcal R_n(e_n)\). The uniform score-moment bound and Lemma~\ref{lem:app-fixed-window-moments} imply \(\|D\Phi_n(\Theta_n)U_n\|_\infty=O_p(1)\) and \(\sqrt{N_n}\|\mathcal R_n(e_n)\|_\infty\le2\sqrt{N_n}\|e_n\|_\infty^2=O_p(N_n^{-1/2})\). Thus \(A_n(K_n)\ge D_n-O_p(1)\). Moreover, \(\|\widehat\Theta_n\|_\infty=O_p(1)\), so
\[
\E^*\{A_n^*(K_n)\}^2\le\|D\Phi_n(\widehat\Theta_n)\|_{\infty\to\infty}^2\E^*\|U_n^*\|_2^2\le16\|\widehat\Theta_n\|_\infty^2\E^*\|U_n^*\|_2^2=O_p(1).
\]
Conditional Markov's inequality gives \(\widehat c_{n,1-\alpha}^{\mathrm{curv}}(K_n)\le[\E^*\{A_n^*(K_n)\}^2/\alpha]^{1/2}=O_p(1)\). Consequently, \(\mathbb{P}\{A_n(K_n)\le\widehat c_{n,1-\alpha}^{\mathrm{curv}}(K_n)\}\le\mathbb{P}\{\|D\Phi_n(\Theta_n)U_n\|_\infty+2\sqrt{N_n}\|e_n\|_\infty^2+\widehat c_{n,1-\alpha}^{\mathrm{curv}}(K_n)\ge D_n\}\to0\), proving consistency.
\end{proof}

\section{Proofs of Lemmas}
\label{sec:supp-lemma-proofs}

\subsection{Proof of Lemma~\ref{lem:dependence-transfer}}

\begin{proof}
Put \(u=q(1+\gamma)\). The growth factor in Assumption~\ref{asm:data-score}(ii), evaluated at the original and coupled observations, has the form \(A^\gamma\), where \(\|A\|_u\le C\) by stationarity, equality of the coupled marginal laws, and Assumption~\ref{asm:data-score}(iii). For \(\gamma>0\), H\"older's inequality gives \(\|A^\gamma D\|_q\le\|A\|_u^\gamma\|D\|_u\); for \(\gamma=0\), the same conclusion holds directly. Take \(D=d(X_{t-k},X_{t-k}^{*(t,j)})+d(X_t,X_t^{*(t,j)})\). Causality makes its first term zero for \(j<k\), and otherwise its two \(L_u\)-norms are \(\delta_{j-k,u}^X\) and \(\delta_{j,u}^X\). This proves part~(i).

Taking \((x_2,y_2)=(x_0,x_0)\) in the score condition gives \(|h(X_{t-k},X_t)|\le |h(x_0,x_0)|+C\{1+d(X_{t-k},x_0)+d(X_t,x_0)\}^{1+\gamma}\). Its \(L_q\)-norm is uniformly bounded by the \(u\)-moment condition. Centering preserves this bound, proving part~(ii).

\end{proof}

\subsection{Proof of Lemma~\ref{lem:app-partial-sum-moment}}

\begin{proof}
Fix $1\le k\le K_n$. The process $\{\widetilde W_{t,n,k}\}_{t\in\mathbb Z}$ is stationary, centered, and causal. By \citet[Theorem~1(i)--(ii)]{wu2005nonlinear} and Lemma~\ref{lem:dependence-transfer}(i) with $t=j$,
\begin{equation}
\sum_{j=0}^{\infty} \|\mathcal P_0\widetilde W_{j,n,k}\|_q \le \sum_{j=0}^{\infty} \| \widetilde W_{j,n,k} - \widetilde W_{j,n,k}^{*(j)} \|_q \le C \sum_{j=0}^{\infty}\delta_{j,q(1+\gamma)}^X ,
\label{eq:app-projection-sum}
\end{equation}
which is finite by Assumption~\ref{asm:data-score} and does not depend on $n$ or $k$. Let $I\subset\mathbb Z$ be a finite interval. By stationarity, $\sum_{t\in I}\widetilde W_{t,n,k}$ has the same distribution as $\sum_{t=1}^{|I|}\widetilde W_{t,n,k}$. Since $q\ge4$, \citet[Theorem~2(i)]{wu2005nonlinear} yields a constant $B_q$ depending only on $q$ such that
\[
\left\| \sum_{t\in I}\widetilde W_{t,n,k} \right\|_q = \left\| \sum_{t=1}^{|I|}\widetilde W_{t,n,k} \right\|_q \le \left\| \max_{1\le m\le|I|} \left| \sum_{t=1}^{m}\widetilde W_{t,n,k} \right| \right\|_q \le B_q|I|^{1/2} \sum_{j=0}^{\infty} \|\mathcal P_0\widetilde W_{j,n,k}\|_q .
\]
Combining this with \eqref{eq:app-projection-sum} proves both the interval and maximal bounds with a constant independent of \(n\), \(k\), and the interval. Stationarity shifts the initial index to any \(a\), and \(h(X_t,X_{t+k})-\theta_k=\widetilde W_{t+k,n,k}\) gives the forward-score statements.
\end{proof}

\subsection{Proof of Lemma~\ref{lem:app-covariance-summability}}

\begin{proof}
\emph{Step 1: Projection bounds.}
For $j\ge0$, write $a_{j,n,k}=\|\mathcal P_0\widetilde W_{j,n,k}\|_2$, and set $a_{j,n,k}=0$ for $j<0$. The projection--coupling inequality and Lemma~\ref{lem:dependence-transfer}(i) give
\[
a_{j,n,k}\le C\left\{\delta_{j,q(1+\gamma)}^X+\1\{j\ge k\}\delta_{j-k,q(1+\gamma)}^X\right\}.
\]
Since $\beta>2$, summing this bound yields
\begin{equation}
\sum_{j\ge0}a_{j,n,k}\le C,\qquad \sum_{j\ge0}j\,a_{j,n,k}\le C(1+k),
\label{eq:app-projection-sum-L2}
\end{equation}
uniformly in $n$ and $k$.

\emph{Step 2: Covariance sums.}
The remote past $\mathcal F_{-\infty}$ is trivial by Kolmogorov's zero--one law \citep[Theorem~2.5.3]{durrett2019probability}, so the centered scores admit their $L^2$ martingale projection expansions. For $\widetilde\gamma_{n,k\ell}(r)=\Cov(\widetilde W_{0,n,k},\widetilde W_{r,n,\ell})$, orthogonality, stationarity, and Cauchy--Schwarz give
\[
|\widetilde\gamma_{n,k\ell}(r)|\le\sum_{j\in\mathbb Z}a_{j,n,k}a_{j+r,n,\ell},\qquad r\in\mathbb Z.
\]
Tonelli's theorem and \eqref{eq:app-projection-sum-L2} imply
\[
\begin{aligned}
\sum_{r\in\mathbb Z}|\widetilde\gamma_{n,k\ell}(r)|&\le\sum_{j,u\ge0}a_{j,n,k}a_{u,n,\ell}\le C,\\
\sum_{r\in\mathbb Z}|r|\,|\widetilde\gamma_{n,k\ell}(r)|&\le\sum_{j,u\ge0}(j+u)a_{j,n,k}a_{u,n,\ell}\le C(1+k+\ell)\le CK_n.
\end{aligned}
\]

\emph{Step 3: Forward scores.}
For $W_{t,n}=H_{t,n}-\Theta_n$, the identity $W_{t,n,k}=\widetilde W_{t+k,n,k}$ gives
\[
\Cov(W_{0,n,k},W_{r,n,\ell})=\widetilde\gamma_{n,k\ell}(r+\ell-k).
\]
Changing variables to $v=r+\ell-k$ preserves the absolute covariance sum. Since $|r|\le|v|+|k-\ell|$, the weighted sum is bounded by
\[
\sum_{v\in\mathbb Z}\bigl(|v|+|k-\ell|\bigr)|\widetilde\gamma_{n,k\ell}(v)|\le CK_n.
\]
Thus $\Sigma_n$ is well defined and $\|\Sigma_n\|_{\max}\le C$. For fixed $n$, absolute covariance summability also gives
\[
L^{-1}\Cov\left(\sum_{t=1}^LW_{t,n}\right)\longrightarrow\Sigma_n\qquad\text{as }L\to\infty.
\]
Each matrix on the left is positive semidefinite, so the same holds for $\Sigma_n$.
\end{proof}

\subsection{Proof of Lemma~\ref{lem:app-blocking-mdep}}

\begin{proof}
Write $K=K_n$, $W_t=W_{t,n}$, and $V_t=V_{t,n}^{(m)}$.

\emph{Step 1: Finite dependence.}
Each $V_t$ is the same measurable function of $(\varepsilon_{t-m+1},\ldots,\varepsilon_{t+K})$. The i.i.d.\ innovations and $\E W_t=0$ give stationarity and centering, while disjoint innovation windows give independence for index sets separated by at least $K+m$. This proves part~(i).

\emph{Step 2: Approximation error.}
Fix a coordinate $k$. For $l\ge-K$, let $\mathcal G_{t,l}=\sigma(\varepsilon_{t-l},\ldots,\varepsilon_{t+K})$, with $\mathcal G_{t,-K-1}$ trivial, and define $D_{t,l}=\E(W_{t,k}\mid\mathcal G_{t,l})-\E(W_{t,k}\mid\mathcal G_{t,l-1})$. Martingale convergence gives $W_{t,k}-V_{t,k}=\sum_{l\ge m}D_{t,l}$ in $L^q$.

Write $\delta_j=\delta_{j,q(1+\gamma)}^X$ for $j\ge0$ and $\delta_j=0$ otherwise. Replacing $\varepsilon_{t-l}$ by an independent copy, conditional Jensen's inequality and Lemma~\ref{lem:dependence-transfer}(i) give
\[
\|D_{t,l}\|_q\le C(\delta_l+\delta_{l+k}),\qquad \sum_{l\ge m}(\delta_l+\delta_{l+k})\le Cm^{1-\beta}.
\]
For fixed $l$, $D_{t,l}$ is measurable with respect to $\sigma(\varepsilon_{t-l},\varepsilon_{t-l+1},\ldots)$ and satisfies
\[
\E(D_{t,l}\mid\sigma(\varepsilon_{t-l+1},\varepsilon_{t-l+2},\ldots))=0.
\]
Reversing time therefore gives martingale differences. Each forward partial sum is the difference between the terminal reversed sum and a reversed partial sum, so Doob's and Burkholder's inequalities yield
\[
\left\|\max_{1\le u\le L}\left|\sum_{t=a+1}^{a+u}D_{t,l}\right|\right\|_q\le C\left(\sum_{t=a+1}^{a+L}\|D_{t,l}\|_q^2\right)^{1/2}\le C\sqrt L(\delta_l+\delta_{l+k}).
\]
Summing over $l\ge m$ by Minkowski's inequality proves part~(ii). Part~(iii) follows from part~(ii) and the forward-score maximal bound in Lemma~\ref{lem:app-partial-sum-moment}.

\emph{Step 3: Covariance approximation.}
Parts~(ii)--(iii) and Lemma~\ref{lem:app-partial-sum-moment} bound the $L^2$ norms of the length-$r$ sums of $W_{t,k}$ and $V_{t,k}$ by $C\sqrt r$, and their difference by $C\sqrt r\,m^{1-\beta}$. Expanding the covariance difference and applying Cauchy--Schwarz gives
\[
\left\|\frac1r\Cov\left(\sum_{t=1}^rV_t\right)-\frac1r\Cov\left(\sum_{t=1}^rW_t\right)\right\|_{\max}\le Cm^{1-\beta}.
\]
Stationarity and Lemma~\ref{lem:app-covariance-summability} also give
\[
\left\|\frac1r\Cov\left(\sum_{t=1}^rW_t\right)-\Sigma_n\right\|_{\max}\le\frac1r\max_{k,\ell\le K}\sum_{u\in\mathbb Z}|u|\,|\Cov(W_{0,k},W_{u,\ell})|\le\frac{CK}{r}.
\]
Combining the two bounds proves part~(iv).
\end{proof}
\subsection{Proof of Lemma~\ref{lem:app-blocking-fluctuation}}

\begin{proof}
Write $N=N_n$, $K=K_n$, $m=m_n$, $r=r_n$, $M=K+m$, $P=r+M$, and $\ell=\ell_n$. Use the finite-memory scores $V_t=V_{t,n}^{(m)}$ of Lemma~\ref{lem:app-blocking-mdep}. Put $J=\lfloor N/P\rfloor$ and define $Y_j=\sum_{t=(j-1)P+1}^{(j-1)P+r}V_t$ and $Y'_j=\sum_{t=(j-1)P+r+1}^{jP}V_t$. For all sufficiently large $n$, $r\ge M$, so the $Y_j$ are independent, as are the $Y'_j$.

Lemma~\ref{lem:app-blocking-mdep}(ii) bounds the expected replacement error, after normalization by $\sqrt N$, by $CK^{1/q}m^{1-\beta}=o(1)$. The partial-sum maximum of $V_t$ is bounded by
\[
\max_{i\le J}\left\|\sum_{j\le i}Y_j\right\|_\infty+\max_{i\le J}\left\|\sum_{j\le i}Y'_j\right\|_\infty+\max_{0\le i\le J}\max_{0\le w\le P}\left\|\sum_{t=iP+1}^{iP+w}V_t\right\|_\infty.
\]
Doob's and Burkholder's inequalities and Lemma~\ref{lem:app-blocking-mdep}(iii) bound the expectations of the second and third terms, divided by $\sqrt N$, by $CK^{1/q}(M/r)^{1/2}$ and $CK^{1/q}(P/N)^{1/2-1/q}$, respectively. Both tend to zero under Assumption~\ref{asm:block-complexity}(i).

For the first term, apply \citet[Lemma~E.1]{chernozhukov2017central} to independent vectors with coordinates $\widetilde X_{j,(i,k)}=\1\{j\le i\}Y_{j,k}$. Their maximal summed coordinate variance is at most $CJr$, and
\[
\E\max_{j,i,k}|\widetilde X_{j,(i,k)}|^2\le\left(\sum_{j,k}\E|Y_{j,k}|^q\right)^{2/q}\le C(JK)^{2/q}r.
\]
Since $J\le N/P$ and $\log(JK)\le\ell$, the cited inequality gives
\[
N^{-1/2}\E\max_{i\le J}\left\|\sum_{j\le i}Y_j\right\|_\infty\le C\left(\frac{Jr}{N}\right)^{1/2}\sqrt\ell+C(JK)^{1/q}\left(\frac rN\right)^{1/2}\ell\le C\sqrt\ell+CK^{1/q}\left(\frac PN\right)^{1/2-1/q}\ell.
\]
The last term tends to zero by the third blocking condition. Combining these bounds proves the result.
\end{proof}

\subsection{Proof of Lemma~\ref{lem:app-anti-concentration}}

\begin{proof}
For part~(i), let \(\bar Z_n=(Z_n^\top,-Z_n^\top)^\top\), so \(M_n=\max_{j\le2p_n}\bar Z_{n,j}\). Nazarov's inequality \citep[Lemma~A.1]{chernozhukov2017central}, applied to \(\bar Z_n\), gives \(\sup_u\mathbb{P}(u-t<M_n\le u+t)\le Ct\sqrt{\log(2p_n)}\), proving part~(i). For part~(ii), adjoining negatives leaves the max-norm covariance discrepancy unchanged. The Gaussian comparison bound in \citet[Theorem~2]{chernozhukov2015comparison} therefore applies to the signed vectors and gives the stated expression for \(\Delta_n>0\). When \(\Delta_n=0\), the two centered Gaussian laws coincide.
\end{proof}

\subsection{Proof of Lemma~\ref{lem:app-quantile-calibration}}

\begin{proof}
After removing zero-variance coordinates, $\|Z_n\|_\infty$ is a finite maximum of absolute values of continuous random variables, so $F_n$ is continuous. For a fixed sufficiently small \(\eta>0\), let \(q_n^-\) and \(q_n^+\) be the deterministic \(F_n\)-quantiles at levels \(1-\alpha-2\eta\) and \(1-\alpha+2\eta\). On the event that the bootstrap distribution differs from \(F_n\) by at most \(\eta\), the definition of the quantile gives \(q_n^-\le c_n^*\le q_n^+\). This event has probability tending to one. Continuity of $F_n$ gives \(F_n(q_n^\pm)=1-\alpha\pm2\eta\), so the sampling approximation bounds \(\mathbb{P}(T_n>c_n^*)\) between \(\alpha-2\eta-o(1)\) and \(\alpha+2\eta+o(1)\). Letting \(\eta\downarrow0\) proves the claim.
\end{proof}

\subsection{Proof of Lemma~\ref{lem:app-block-covariance}}

\begin{proof}
Write $W_{t,n}=H_{t,n}-\Theta_n$, $J_n=\lfloor N_n/b_n\rfloor$, and $S_{\ell,n}=\sum_{t=(\ell-1)b_n+1}^{\ell b_n}W_{t,n}$. Let $C_n^0$ be the kernel obtained from $\widehat C_n$ by replacing $\widehat\Theta_n$ with $\Theta_n$.

\emph{Step 1: Complete blocks.}
For $s\le s'$, put $m=\lfloor N_ns\rfloor$, $m'=\lfloor N_ns'\rfloor$, and $J=\lfloor m/b_n\rfloor$. Then
\[
C_n^0(s,s')=\frac1{N_n}\sum_{\ell=1}^JS_{\ell,n}S_{\ell,n}^\top+R_n(s,s'),
\]
where
\[
R_n(s,s')=\frac1{N_n}\left(\sum_{t=Jb_n+1}^{m}W_{t,n}\right)\left(\sum_{t=Jb_n+1}^{m'\wedge((J+1)b_n)}W_{t,n}\right)^\top.
\]
Stationarity and the definition of $\mathfrak b_n$ give
\[
\max_{0\le J\le J_n}\left\|\frac1{N_n}\sum_{\ell=1}^J\{\E(S_{\ell,n}S_{\ell,n}^\top)-b_n\Sigma_n\}\right\|_{\max}\le\mathfrak b_n.
\]

Set $p=q/2\ge2$ and $V_{\ell,n}^{kj}=b_n^{-1}\{S_{\ell,n,k}S_{\ell,n,j}-\E(S_{\ell,n,k}S_{\ell,n,j})\}$. Lemma~\ref{lem:app-partial-sum-moment} and Hölder's inequality give $\|S_{\ell,n,k}\|_q\le C\sqrt{b_n}$ and $\|V_{\ell,n}^{kj}\|_p\le C$. Moreover,
\begin{equation}
\max_{0\le J\le J_n}\left\|\frac1{N_n}\sum_{\ell=1}^J\{S_{\ell,n}S_{\ell,n}^\top-\E(S_{\ell,n}S_{\ell,n}^\top)\}\right\|_{\max}=\frac{b_n}{N_n}\max_{k,j\le K_n}\max_{0\le J\le J_n}\left|\sum_{\ell=1}^JV_{\ell,n}^{kj}\right|.
\label{eq:max-V}
\end{equation}

Group the innovations as $\eta_\ell=(\varepsilon_{K_n+(\ell-1)b_n+1},\ldots,\varepsilon_{K_n+\ell b_n})$. Then $V_{\ell,n}^{kj}$ is a stationary causal function of these independent groups. Replace $\eta_0$ by an independent copy and mark the resulting variables with a superscript $*$. For $v\in\{0,k,j\}$, define $c_{v,n}=\lfloor(K_n-v)/b_n\rfloor$ and $\mathcal I_n^{kj}=\bigcup_{v\in\{0,k,j\}}\{c_{v,n},c_{v,n}+1,c_{v,n}+2\}$. Thus $|\mathcal I_n^{kj}|\le9$.

The observation window with offset $v$ is unaffected when $\ell\le c_{v,n}-1$. When $\ell\ge c_{v,n}+3$, it lies at least $(\ell-c_{v,n}-2)b_n$ time steps after the replaced innovations. Replacing these innovations one at a time and using Lemma~\ref{lem:dependence-transfer} therefore gives, for $\ell\notin\mathcal I_n^{kj}$,
\[
\|S_{\ell,n,k}-S_{\ell,n,k}^*\|_q\le Cb_n^{2-\beta}\sum_{\substack{v\in\{0,k\}\\\ell\ge c_{v,n}+3}}(\ell-c_{v,n}-2)^{-\beta},
\]
and similarly for coordinate $j$. For $\ell\in\mathcal I_n^{kj}$, the same difference is bounded by $C\sqrt{b_n}$. Expanding the product difference and applying Hölder's inequality yields
\[
\|V_{\ell,n}^{kj}-V_{\ell,n}^{kj,*}\|_p\le C\1\{\ell\in\mathcal I_n^{kj}\}+Cb_n^{3/2-\beta}\sum_{\substack{v\in\{0,k,j\}\\\ell\ge c_{v,n}+3}}(\ell-c_{v,n}-2)^{-\beta}.
\]
Since $\beta>2$ and $|\mathcal I_n^{kj}|\le9$, we have $\sum_{\ell\ge0}\|V_{\ell,n}^{kj}-V_{\ell,n}^{kj,*}\|_p\le C$ uniformly in $n,k,j$. Applying the projection and martingale maximal argument from Lemma~\ref{lem:app-partial-sum-moment} to the grouped innovations, with exponent $p\ge2$, gives
\[
\left\|\max_{0\le J\le J_n}\left|\sum_{\ell=1}^JV_{\ell,n}^{kj}\right|\right\|_p\le C\sqrt{J_n}.
\]
Taking the maximum over $K_n^2$ coordinate pairs in \eqref{eq:max-V} bounds the centered complete-block contribution by
\[
O_p\left\{K_n^{2/p}\frac{b_n}{N_n}\sqrt{J_n}\right\}=O_p\left\{K_n^{4/q}\left(\frac{b_n}{N_n}\right)^{1/2}\right\}.
\]

\emph{Step 2: Partial blocks.}
Define $M_{\ell,n,k}=\max_{0\le u\le b_n}|\sum_{t=(\ell-1)b_n+1}^{(\ell-1)b_n+u}W_{t,n,k}|$, extending the final block to length $b_n$. Lemma~\ref{lem:app-partial-sum-moment} gives $\|M_{\ell,n,k}\|_q\le C\sqrt{b_n}$, and hence
\[
\sup_{s\le s'}\|R_n(s,s')\|_{\max}\le\frac1{N_n}\max_{\ell,k}M_{\ell,n,k}^2=O_p\left\{K_n^{2/q}\left(\frac{b_n}{N_n}\right)^{1-2/q}\right\}.
\]
The unmatched Gaussian contribution is at most $Cb_n/N_n$, since $0\le m-Jb_n<b_n$ and $\|\Sigma_n\|_{\max}\le C$. Both terms are absorbed by the bound in Step~1 because $q\ge4$. Transposition handles $s>s'$, so
\[
\sup_{s,s'\in[0,1]}\|C_n^0(s,s')-C_n^G(s,s')\|_{\max}=O_p(\rho_n).
\]

\emph{Step 3: Sample centering.}
Let $e_n=\widehat\Theta_n-\Theta_n$, $A_{\ell,n}(s)=\sum_{t\in I_{\ell,n}:\,t\le\lfloor N_ns\rfloor}W_{t,n}$, and $c_{\ell,n}(s)=|\{t\in I_{\ell,n}:t\le\lfloor N_ns\rfloor\}|$. The sample-centered block sum is $A_{\ell,n}(s)-c_{\ell,n}(s)e_n$. Lemma~\ref{lem:app-partial-sum-moment} and Minkowski's inequality give
\[
\|e_n\|_\infty=O_p(K_n^{1/q}N_n^{-1/2}),\qquad \max_k\frac1{N_n}\sum_{\ell=1}^{L_n}M_{\ell,n,k}^2=O_p(K_n^{2/q}).
\]
Since $|A_{\ell,n,k}(s)|\le M_{\ell,n,k}$ and $\sum_\ell c_{\ell,n}(s)^2\le b_nN_n$, expansion of the covariance products and Cauchy--Schwarz yield
\[
\begin{aligned}
\sup_{s,s'}\|\widehat C_n(s,s')-C_n^0(s,s')\|_{\max}&\le2\|e_n\|_\infty\sqrt{b_n}\left(\max_k\frac1{N_n}\sum_\ell M_{\ell,n,k}^2\right)^{1/2}+b_n\|e_n\|_\infty^2\\
&=O_p\left\{K_n^{2/q}\sqrt{\frac{b_n}{N_n}}\right\}.
\end{aligned}
\]
Combining this with Step~2 proves part~(i).

\emph{Step 4: Transformations and comparison.}
The terminal and bridge covariance errors satisfy
\[
\begin{aligned}
r_{n,A}^{\mathrm{boot}}&\le\|\mathsf A_n\|_{\infty\to\infty}^2\|\widehat C_n(1,1)-C_n^G(1,1)\|_{\max}=O_p(\rho_n),\\
r_{n,\mathrm{br}}^{\mathrm{boot}}&\le4\sup_{s,s'\in[0,1]}\|\widehat C_n(s,s')-C_n^G(s,s')\|_{\max}=O_p(\rho_n),
\end{aligned}
\]
which proves part~(ii).

For part~(iii), let $r_n$ denote either covariance error and let $d_n$ be the corresponding dimension $d_{A,n}$ or $p_n$. Assumption~\ref{asm:block-complexity}(ii) and $d_n\le d_n^{\max}$ give
\[
r_n=O_p(\rho_n)=o_p(1),\qquad r_n\{1\vee\log(2d_n)\}^2=o_p(1).
\]
On $0<r_n<1$,
\[
r_n\{1\vee\log(2d_n/r_n)\}^2\le2r_n\{1\vee\log(2d_n)\}^2+2r_n\log^2(1/r_n)=o_p(1),
\]
because $x\log^2(1/x)\to0$ as $x\downarrow0$. Taking cube roots, with the stated convention at zero, proves part~(iii).
\end{proof}
\subsection{Proof of Lemma~\ref{lem:app-fixed-window-moments}}

\begin{proof}
Apply Lemma~\ref{lem:app-partial-sum-moment} coordinatewise to the forward score sequence to obtain the first bound and \(\E\|\widehat\Theta_n-\Theta_n\|_2^2\le C/N_n\). Conditional independence of the multipliers gives \(\E^*\|U_n^*\|_2^2=N_n^{-1}\sum_{\ell=1}^{L_n}\|\sum_{t\in I_{\ell,n}}(H_{t,n}-\widehat\Theta_n)\|_2^2\). Replacing \(\widehat\Theta_n\) by \(\Theta_n\) and using \(\|x-y\|_2^2\le2\|x\|_2^2+2\|y\|_2^2\), its expectation is at most \(C+C N_n^{-2}\sum_\ell|I_{\ell,n}|^2\le C(1+b_n/N_n)\). Markov's inequality gives the sampling orders, and its conditional version gives the bootstrap tightness.
\end{proof}

\subsection{Proof of Lemma~\ref{lem:app-cp-drift}}

\begin{proof}
Replace the actual mean score at anchor \(t\) by \(\Theta_n^{(1)}\) for \(t\le m_0\) and by \(\Theta_n^{(2)}\) for \(t>m_0\). For coordinate \(k\), these replacements differ from the actual means only for at most \(k\) pairs crossing the change. Their contribution to the centered unnormalized cumulative sum is bounded by \(Ck\), uniformly in the endpoint. Let \(\lambda=\lfloor N_ns\rfloor/N_n\) and \(\tau=m_0/N_n\). For the replaced sequence, the mean bridge is exactly \(\sqrt{N_n}\{\min(\lambda,\tau)-\lambda\tau\}(\Theta_n^{(1)}-\Theta_n^{(2)})\). Since \(|\lambda-s|\le1/N_n\), \(|\tau-\tau_0|\le1/N_n\), and the regime means are bounded, replacing \(\lambda,\tau\) by \(s,\tau_0\) adds \(O(N_n^{-1/2})\). The resulting expression is \(\sqrt{N_n}\mathcal D_n(s)\) in \eqref{eq:cp-deterministic-drift}. Maximizing the crossing error over \(k\le K_n\) proves the result.
\end{proof}

\subsection{Proof of Lemma~\ref{lem:curvature-taylor}}

\begin{proof}
Expanding \((\theta_k+e_k)(\theta_{k+2}+e_{k+2})-(\theta_{k+1}+e_{k+1})^2\) gives the population coordinate, the linear term \(\theta_{k+2}e_k-2\theta_{k+1}e_{k+1}+\theta_ke_{k+2}\), and the remainder \(e_ke_{k+2}-e_{k+1}^2\). The remainder bound follows by the triangle inequality. Each Jacobian row has entries \(\theta_{k+2},-2\theta_{k+1},\theta_k\) in its three relevant columns, so its absolute row sum is at most \(4\|\theta\|_\infty\), and the row sum of the Jacobian difference is at most \(4\|e\|_\infty\).
\end{proof}

\clearpage
\phantomsection
\addcontentsline{toc}{section}{References}
\bibliographystyle{apalike}
\bibliography{ref}
\end{document}